\documentclass[a4paper,11p]{article}
\pdfoutput=1 % if your are submitting a pdflatex 
\usepackage{jcappub} % for details on the use of the package, 
\usepackage{comment}
\usepackage{color}
\usepackage[normalem]{ulem}
\usepackage{mathrsfs}

\graphicspath{figures}

\newcommand{\de}{\mathrm{d}}
\renewcommand{\vec}{\mathbf}

\newcommand{\be}{\begin{equation}}
\newcommand{\ee}{\end{equation}}
\newcommand{\bea}{\begin{eqnarray}}
\newcommand{\eea}{\end{eqnarray}}
\newcommand{\bdm}{\begin{displaymath}}
\newcommand{\edm}{\end{displaymath}}
\newcommand{\nn}{\nonumber}

\newcommand{\xv}{\mathbf{x}}

\newcommand{\kv}{\mathbf{k}}

\newcommand{\qv}{\mathbf{q}}

\newcommand{\Pl}{P_{\rm L}}
\newcommand{\del}{\delta}
\newcommand{\G}{\mathcal G}

\newcommand{\kmax}{k_{\rm max}}
\newcommand{\kmaxP}{k_{\rm max,P}}
\newcommand{\kmaxB}{k_{\rm max,B}}
\newcommand{\Pnw}{P_{\rm nw}}
\newcommand{\Pw}{P_{\rm w}}
\newcommand{\PL}{P_{\rm L}}
\newcommand{\PLO}{P_{\rm LO}}
\newcommand{\PNLO}{P_{\rm NLO}}

\newcommand{\bG}{b_{\mathcal G_2}}
\newcommand{\bgamma}{b_{\Gamma_3}}
\newcommand{\bk}{b_{\nabla^2 \delta}}
\newcommand{\keff}{k_{\rm eff}}
\newcommand{\keffl}{k_{{\rm eff}, l}}
\newcommand{\keffm}{k_{{\rm eff}, m}}

\newcommand{\Kt}{\widetilde{K}}

\newcommand{\C}{{\mathbb C}}
\newcommand{\Ce}{{\widetilde{\mathbb C}}}

\newcommand{\ie}{{\em i.e.}~}
\newcommand{\eg}{{\em e.g.}}
\newcommand{\Ms}{\, h^{-1} \, \mathrm{M}_\odot}
\newcommand{\Mpc}{\, h^{-1} \, {\rm Mpc}}

\newcommand{\icMpc}{\, h^{3} \, {\rm Mpc}^{-3}}
\newcommand{\Gpc}{\, h^{-1} \, {\rm Gpc}}
\newcommand{\cGpc}{\, h^{-3} \, {\rm Gpc}^3}
\newcommand{\kMpc}{\, h \, {\rm Mpc}^{-1}}
\newcommand{\sMpc}{\, h^{-2} \, {\rm Mpc}^{2}}

\newcommand{\pin}{{\sc{Pinocchio}}}
\newcommand{\mine}{{\sc{Minerva}}}

\defcitealias{OddoEtal2020}{Paper I}
\newcommand{\OddoEtal}{\citetalias{OddoEtal2020}}

\title{Cosmological parameters from the likelihood analysis of the galaxy power spectrum and bispectrum in real space}

\author[a,b,1]{Andrea Oddo,\note{Corresponding author.}}
\author[b,c,d,e]{Federico Rizzo,}
\author[b,d,e]{Emiliano Sefusatti,}
\author[f]{Cristiano Porciani,}
\author[c,e,d,b]{Pierluigi Monaco}

\affiliation[a]{SISSA - International School for Advanced Studies, Via Bonomea 265, 34136 Trieste, Italy}
\affiliation[b]{Institute for Fundamental Physics of the Universe, Via Beirut 2, 34151 Trieste, Italy}
\affiliation[c]{Dipartimento di Fisica, Sezione di Astronomia, Universit\`a di Trieste, via Tiepolo 11, 34143 Trieste, Italy}
\affiliation[d]{Istituto Nazionale di Fisica Nucleare, Sezione di Trieste, via Valerio 2, 34127 Trieste, Italy}
\affiliation[e]{Istituto Nazionale di Astrofisica, Osservatorio Astronomico di Trieste, via Tiepolo 11, 34143 Trieste, Italy}
\affiliation[f]{Argelander Institut f\"ur Astronomie der Universit\"at Bonn, Auf dem H\"ugel 71, 53121 Bonn, Germany}

\emailAdd{andrea.oddo@sissa.it}
\emailAdd{federico.rizzo@inaf.it}
\emailAdd{emiliano.sefusatti@inaf.it}
\emailAdd{porciani@astro.uni-bonn.de}
\emailAdd{pierluigi.monaco@inaf.it}

\abstract{ 
We present a joint likelihood analysis of the halo power spectrum and bispectrum in real space. We take advantage of a large set of numerical simulations and of an even larger set of halo mock catalogs to provide a robust estimate of the covariance properties. We derive constraints on bias and cosmological parameters assuming a theoretical model from perturbation theory at one-loop for the power spectrum and tree-level for the bispectrum. By means of the Deviance Information Criterion, we select a reference bias model dependent on seven parameters that can describe the data up to $\kmaxP=0.3\kMpc$ for the power spectrum and $\kmaxB=0.09\kMpc$ for the bispectrum at redshift $z=1$. This model is able to accurately recover three selected cosmological parameters even for the rather extreme total simulation volume of 1000$\cGpc$. With the same tools, we study how relations among bias parameters can improve the fit while reducing the parameter space. In addition, we compare common approximations to the covariance matrix against the full covariance estimated from the mocks, and quantify the (non-negligible) effect of ignoring the cross-covariance between the two statistics. Finally, we explore different selection criteria for the triangular configurations to include in the analysis, showing that excluding nearly equilateral triangles rather than simply imposing a fixed maximum $\kmaxB$ on all triangle sides can lead to a better exploitation of the information contained in the bispectrum.
}

\begin{document}

\maketitle

%%%%%%%%%%%%%%%%%%%%%%%%%%%%%%%%%%%%%%%%%%%%%%%%
%%%%%%%%%%%%%%%%%%%%%%%%%%%%%%%%%%%%%%%%%%%%%%%%
%%%%%%%%%%%%%%%%%%%%%%%%%%%%%%%%%%%%%%%%%%%%%%%%
\section{Introduction}

In the next few years, galaxy redshift surveys will probe the large-scale structure (LSS) over very big volumes, with the foremost objective of determining the origin of the accelerated expansion of the Universe \cite{DawsonEtal2016, LaureijsEtal2011, LeviEtal2013}. 

The two-point correlation function (2PCF) in configuration space and the galaxy power spectrum in Fourier space constitute the main probes of the large-scale galaxy distribution, as they contain the bulk of the information on cosmological parameters. Current surveys, however, provide as well precise measurements of higher-order correlation functions \citep{SlepianEtal2017, GilMarinEtal2015, GilMarinEtal2015B, GilMarinEtal2017, PearsonSamushia2018}. The galaxy 3-point correlation function and the galaxy bispectrum are the lowest-order statistics quantifying the non-Gaussianity of the distribution of galaxies in the LSS and are expected to improve or strengthen any analysis solely based on 2-point correlators \citep{SefusattiScoccimarro2005, SefusattiEtal2006,  SongTaruyaOka2015, ByunEtal2017, ChanBlot2017, GagraniSamushia2017, YankelevichPorciani2019, ChudaykinIvanov2019, KamalinejadSlepian2020a, HahnEtal2020, HahnVillaescusaNavarro2021, GualdiVerde2020,  SamushiaSlepianVillaescusaNavarro2021, AgarwalEtal2021, EggemeierEtal2021}, particularly in the context of beyond-the-standard-model cosmologies as non-Gaussian initial conditions \citep{ScoccimarroSefusattiZaldarriaga2004, SefusattiKomatsu2007, Sefusatti2009, SefusattiCrocceDesjacques2012, ScoccimarroEtal2012, TasinatoEtal2014, TellariniEtal2015, TellariniEtal2016, YamauchiYokoyamaTakahashi2017, KaragiannisEtal2018, Barreira2020, MoradinezhadDizgahEtal2021} or modified gravity/dark energy \citep{ShirataEtal2007, YamauchiYokoyamaTashiro2017, BoseTaruya2018, BoseEtal2020, HeinrichDore2020}.

With these motivations, but also as a natural test for the modelling of the galaxy power spectrum, a large number of works sought to improve the theoretical description, in perturbation theory (PT), of the galaxy bispectrum \citep{SmithShethScoccimarro2008, BernaerdeauCrocceScoccimarro2012, RampfWong2012, AssassiEtal2014, GilMarinEtal2014, SaitoEtal2014, BaldaufEtal2015A, AnguloEtal2015, LazanuEtal2016, HashimotoRaseraTaruya2017, IvanovSibiryakov2018, LazanuLiguori2018, DesjacquesJeongSchmidt2018B, NadlerPerkoSenatore2018, SimonovicEtal2018, DeBelsunceSenatore2019,  EggemeierScoccimarroSmith2019, OddoEtal2020, MoradinezhadDizgahEtal2021, EggemeierEtal2021, SteeleBaldauf2021, GualdiGilMarinVerde2021, AlkhanishviliEtal2021}. Only a handful of these works, however, explored the application of such models to a proper likelihood analysis of halo/galaxy catalogs from numerical simulations \citep{SaitoEtal2014, OddoEtal2020, GualdiVerde2020, MoradinezhadDizgahEtal2021, EggemeierEtal2021, GualdiGilMarinVerde2021}. Among these, reference \citep{SaitoEtal2014} considered a test of the one-loop model for the real-space power spectrum and a tree-level model for the bispectrum for several halo catalogs with varying mass and redshifts, where the inclusion of non-local bias corrections was found to be crucial to obtain a coherent description for both statistics.
Reference \cite{EggemeierEtal2021} extended the perturbative description to include one-loop corrections to the tree-level galaxy bispectrum model in real space. Presenting a joint analysis of power spectrum and bispectrum for several halo and galaxy catalogs, the authors showed that the inclusion of one-loop contributions increases significantly the range of validity of the predictions, roughly from $0.17\kMpc$ to $0.3\kMpc$ for an effective volume of $6 \cGpc$, for redshifts between 0 and 1, with no strong dependence on redshift in this range.
The real-space predictions at one-loop for the power spectrum and tree-level for the bispectrum are tested instead in the context of non-Gaussian initial conditions in \citep{MoradinezhadDizgahEtal2021}. 
The recent work presented in \cite{GualdiGilMarinVerde2021}, limited to the analysis of matter statistics, explores the combination of power spectrum, bispectrum, and trispectrum in redshift space, highlighting how the last could give a potentially relevant contribution to constraining cosmological parameters, such as the growth rate and the fluctuations amplitude. The theoretical predictions for the higher-order correlators are, in this case, based on a phenomenological model fitted to N-body simulations, and are expected to extend somehow beyond the reach of the tree-level predictions in PT. 

Reference \citep{OddoEtal2020} focused on the analysis of the halo bispectrum in real space based on a tree-level model, but explored model-selection techniques along with several possible sources of systematic errors in the likelihood analysis. In order to do so, the authors took advantage of a large set of 298 simulations (the \mine{} set, first presented in \cite{GriebEtal2016}) corresponding to a total volume of roughly $1000\cGpc$, paired with an even larger set of 10\,000 halo mock catalogs obtained for the same box and cosmology with the \pin{} code \cite{MonacoTheunsTaffoni2002, MonacoEtal2013, MunariEtal2017}.
In fact, an inherent difficulty in the bispectrum analysis comes from the signal being distributed over a large number of triangular configurations. Making unbiased inferences thus requires the robust estimation of a large covariance matrix, often obtained numerically, using a large set of mock catalogs or N-body simulations \cite{Scoccimarro2000B, SefusattiScoccimarro2005, SefusattiEtal2006, ChanBlot2017, GilMarinEtal2015, GilMarinEtal2017}. Alternatives are represented by different approaches involving the compression of the bispectrum information into a small number of data points \cite{ByunEtal2017, GualdiEtal2018, ByunEtal2021}.

The present work is the natural continuation of \cite{OddoEtal2020} (hereafter \OddoEtal). Here we investigate the consequences of several assumptions that can be made on the joint likelihood analysis of the galaxy power spectrum and bispectrum in real space, and their effects on the constraints of bias and cosmological parameters. We also apply the Bayesian model selection methods considered there to evaluate theoretically or numerically derived relations among the bias parameters that can reduce the parameter space and speed-up the likelihood evaluation. We make use of the same set of \mine{} simulations, allowing, due to the small sample variance, not only an assessment of systematic errors in the theoretical model, but also an evaluation of the effects related to technical details of the likelihood analysis. In addition, we estimate the full covariance, including cross-correlations between power spectrum and bispectrum, from measurements of the \pin{} mocks.

We fit the data with a one-loop PT model for the power spectrum similar to those tested in the blinded challenge presented in \citep{NishimichiEtal2020} (here limited to real space), including a counterterm to account for the dynamics of short-scale perturbations \cite{PueblasScoccimarro2009, BaumannEtal2012, CarrascoHertzbergSenatore2012}, higher-derivative bias \citep{Desjacques2008, McDonaldRoy2009, DesjacquesEtal2010, DesjacquesJeongSchmidt2018} and the infrared (IR) resummation procedure to describe the non-linear evolution of the BAO features \citep{BaldaufEtal2015B, IvanovSibiryakov2018}. Finally, we also consider a possible scale-dependent correction to the constant shot-noise contribution induced by halo exclusion \cite{SmithScoccimarroSheth2007, BaldaufEtal2013}. For the bispectrum we limit the prediction to the tree-level contribution, testing the possible effect of including higher-derivative bias and IR resummation.

We explore the consequences of an improper treatment of discretization effects in the theoretical predictions on the estimate of the parameters, and study different criteria for the selection of the triangular configurations of the bispectrum. In addition, we investigate the effect of common approximations to the covariance matrix. Finally, we present the results of a full likelihood analysis of the combination of the galaxy power spectrum and bispectrum in real space where cosmological parameters are allowed to vary, using the full simulation volume.

This paper is organised as follows. We introduce all numerical data in section \ref{sec:data} and the theoretical model and the likelihood functions in section \ref{sec:model}. Our
results are discussed in section \ref{sec:results}. Finally, we present our conclusions in section \ref{sec:conclusions}.

%%%%%%%%%%%%%%%%%%%%%%%%%%%%%%%%%%%%%%%%%%%%%%%%
%%%%%%%%%%%%%%%%%%%%%%%%%%%%%%%%%%%%%%%%%%%%%%%%
%%%%%%%%%%%%%%%%%%%%%%%%%%%%%%%%%%%%%%%%%%%%%%%%
\section{Data}
\label{sec:data}

\subsection{N-body simulations and measurements}
\label{sec:sims} 

We make use of the same \mine{} set of 298 N-body simulations adopted in \citetalias{OddoEtal2020} and first presented in \cite{GriebEtal2016}. Each run evolves $1000^3$ dark-matter particles in a periodic cubic box of side $L = 1500 \Mpc$, and assumes a flat $\Lambda$CDM cosmology characterised by the Hubble parameter $h=0.695$, the total matter and baryonic relative densities $\Omega_m=0.285$ and $\Omega_b=0.046$, the spectral index $n_s=0.9632$ and $\sigma_8=0.828$. The whole set corresponds to a total volume of roughly $1000 \cGpc$.

We focus on the matter and dark-matter halo distributions at redshift $z=1$, as this value is of particular relevance for upcoming spectroscopic galaxy surveys such as Euclid \cite{LaureijsEtal2011} or DESI (Dark Energy Spectroscopic Instrument) \cite{AghamousaEtal2016} \footnote{Since we consider only one value for the redshift, if not otherwise stated, we drop any time dependence in all equations for compactness.}. The halo population we consider has a minimum mass of $M\simeq 1.12 \times 10^{13} \Ms$, equivalent to 42 dark-matter particles and a mean number density of $\bar n = 2.13 \times 10^{-4} \icMpc$.

Measurements of matter and halo power spectra are obtained from the estimator
\be
\label{eq:PowerEstimator}
    \hat{P}(k) \equiv \frac{1}{N_P(k) \,L^3}\sum_{\qv\in k}\,|\del_{\qv}|^2\,,
\ee
where the Fourier-space density $\delta_\qv$ is the result of fourth-order interpolation and the interlacing technique described in \cite{SefusattiEtal2016}, and $k_f\equiv2\pi/L$ is the fundamental wavenumber of the simulation box\footnote{We adopt the following convention for the discrete Fourier transform
\be
    \delta_\kv  \equiv  \int_V \frac{d^3x}{(2\pi)^3}\, e^{-i\kv\cdot\xv}\,\delta(\xv)\,,
\ee
with the inverse given by the series
\be
    \delta(\xv)  \equiv  k_f^3\, \sum_{\kv}\,e^{i\kv\cdot\xv}\,\delta_\kv\,.
\ee
With this convention, the definitions for the power spectrum and the bispectrum are, respectively
\be
\langle \del_{\kv_1} \del_{\kv_2} \rangle  \equiv  (2 \pi)^3 \del_D (\kv_{12}) P(k_1) \ee
and
\be
\langle \del_{\kv_1} \del_{\kv_2} \del_{\kv_3}\rangle \equiv (2 \pi)^3 \del_D (\kv_{123}) B(\kv_1, \kv_2, \kv_3)\,.
\ee
}.
The sum runs over all discrete vectors $\qv$ in a bin of size $\Delta k$, \ie with $k-\Delta k/2\le|\qv|< k+\Delta k/2$, and $N_P(k)= \sum_{\qv \in k}1$ represents their total number.
The bispectrum estimator is defined as 
\be
\label{eq:BispEstimator}
    \hat{B}(k_1, k_2, k_3) \equiv \frac{1}{N_B(k_1, k_2, k_3)\,L^3}\sum_{\qv_1\in k_1}\sum_{\qv_2\in k_2}\sum_{\qv_3\in k_3} \del_K(\vec q_{123}) \del_{\qv_1}\del_{\qv_2}\del_{\qv_3}\,,
\ee
where $\del_K(\kv)$ denotes the Kronecker delta function (equal to 1 for $\kv=0$ and 0 otherwise) and $\qv_{i_1\dots i_n}\equiv\qv_{i_1}+\dots +\qv_{i_n}$. The normalisation factor
\be
\label{eq:BispEstimatorNorm}
    N_B(k_1, k_2, k_3) \equiv \sum_{\qv_1\in k_1}\sum_{\qv_2\in k_2}\sum_{\qv_3\in k_3} \del_K(\vec q_{123}) \,,
\ee 
corresponds to the total number of wavenumber triplets $(\qv_1,\qv_2,\qv_3)$ forming closed triangles that lie in the ``triangle bin'' defined by the triplet $(k_1,k_2,k_3)$, with the $k_i$'s being the bin centers, and where each bin has a width $\Delta k$. In the rest of the paper we refer to the triplets $(\qv_1,\qv_2,\qv_3)$ formed by wavevectors on the original density grid as ``fundamental triangles'' to distinguish them from the ``triangle bin'' $(k_1,k_2,k_3)$. The implementation for the bispectrum estimator is described in \cite{Scoccimarro2015}.

Following the notation adopted in \citetalias{OddoEtal2020}, we denote with $s=\Delta k/k_f$ the $k$-bin size and with $c$ the center of the first bin, both in units of the fundamental frequency. We focus on the power spectrum and bispectrum measurements with the binning scheme given by $(s, c) = (2, 2.5)$. Imposing a maximum wavenumber $\kmax=0.09\kMpc$, the range of validity (also known as the reach) for the tree-level model found in \OddoEtal{} for the whole \mine{} data-set, an $s=2$ binning leads to a total of 170 triangle bins.

\begin{figure}[!t]
    \centering
    \includegraphics[width=\textwidth]{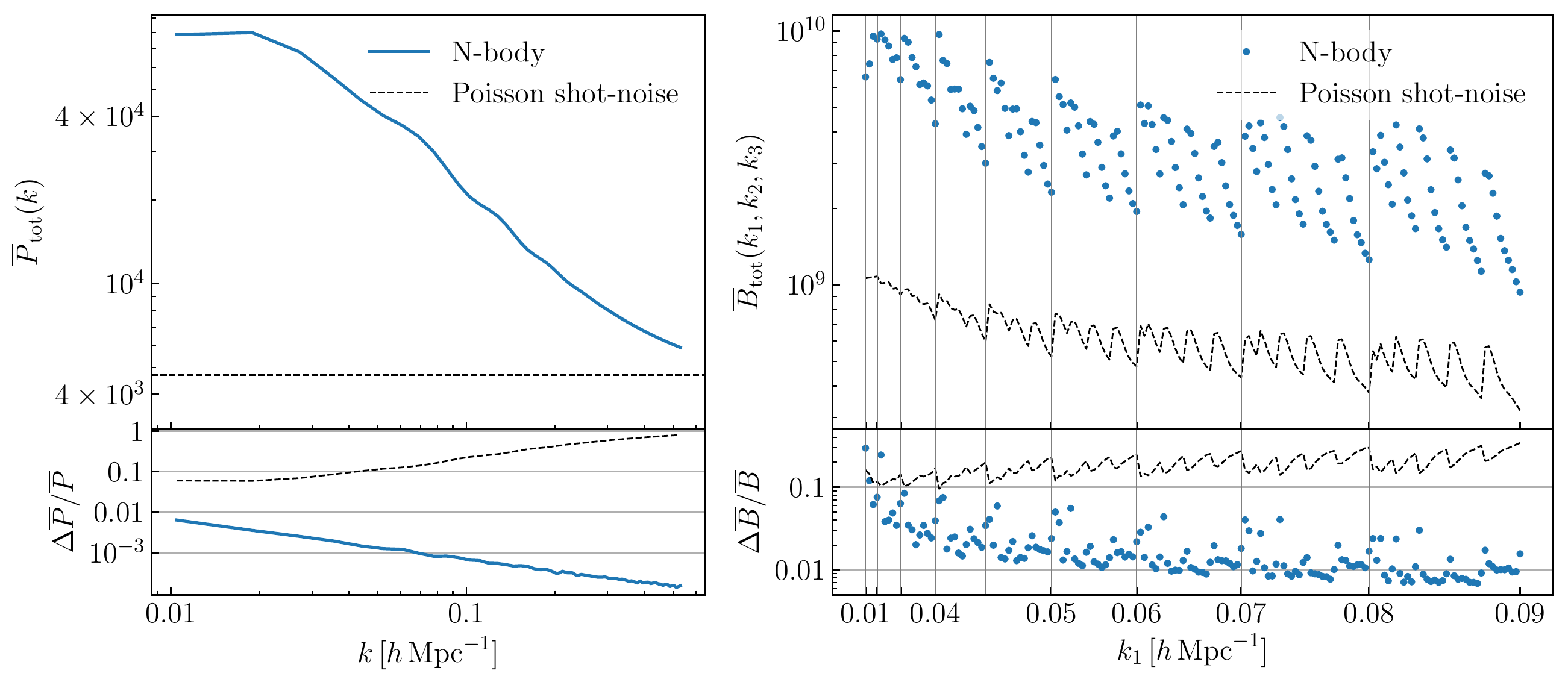}
    \caption{The upper panels show the mean of the measurements of the halo power spectrum (left) and bispectrum (right) extracted from the N-body simulations; the dashed lines represent the Poisson prediction for the shot-noise contributions of the corresponding statistics. The lower panels show the relative standard error on the mean of the power spectrum (left) and the bispectrum (right); the dashed lines show the relative contribution of the Poisson prediction of the shot-noise to the total statistics. In the panels for the bispectrum measurements, we show triangles with increasing values of the sides subject to the constraint $k_1 \ge k_2 \ge k_3$, with gray vertical lines marking the last configuration sharing the corresponding value of $k_1$.}
    \label{fig:measurements}
\end{figure}

The left panels of figure \ref{fig:measurements} show the total halo power spectrum (i.e. including shot-noise), averaged over the full set of 298 N-body simulations (top panel) and the relative error on the mean (bottom panel). Measurements of the cross halo-matter power spectrum (not shown) have also been performed and used for cross-checks and independent estimates of some of the model parameters. The right panels of figure \ref{fig:measurements} show instead the average total halo bispectrum (top panel) and the relative error on the mean (bottom panel). We show all triangle bins with increasing values for the sides subject to the constraint $k_1\ge k_2\ge k_3$. The gray vertical lines and the numbers mark those configurations where the value of $k_1$ changes.

We should notice that the statistical uncertainty on the halo power spectrum is below the percent level even at the largest scales considered, while for the bispectrum it is below the 10 percent level for most of the triangles considered, and in some cases is even sub-percent. It is natural to expect that systematic errors related to the approximations assumed by the N-body solver might then be larger and relevant. We refer the reader to \cite{AlkhanishviliEtal2021} for a study of how systematic errors affect the determination of the reach of PT models on the matter power spectrum and bispectrum. 
Based on that work, we expect systematics to affect to some extent the determination of the bias and cosmological parameters, but we assume their effect to be overall negligible. For this reason, we are not accounting for them in our results.

\subsection{Mock halo catalogs and covariance}

In addition to the \mine{} set of N-body simulations, we also employ a much larger set of 10\,000 mock halo catalogs, generated with the code \pin{} \cite{MonacoTheunsTaffoni2002, MonacoEtal2013, MunariEtal2017} using the same box size and background cosmology as in the N-body simulations. The version of \pin{} used here \cite{MunariEtal2017} relies on a set of criteria, based on ellipsoidal collapse, to group particles into halos, and takes advantage of third-order Lagrangian perturbation theory to displace dark matter halos to their position at $z=1$.
The capability of the \pin{} code to reproduce the covariance properties of the 2-point correlation function, power spectrum, and bispectrum has been established in a series of papers \cite{LippichEtal2019, BlotEtal2019, ColavincenzoEtal2019}.

We use the mock halo catalogs to estimate the covariance matrix for the joint power spectrum and bispectrum measurements. For both the power spectrum and bispectrum, the leading Gaussian contribution to the covariance matrix depends on the amplitude of the total halo power spectrum. Therefore, we require the total halo power spectrum of the mock catalogs to match the one of the N-body simulations at large scales, by adjusting the mass threshold in the \pin{} mocks. This is done in order to minimize the systematic differences between the covariance matrices extracted from the mocks and the ones from the N-body simulations, and thus to allow for an assessment of the goodness of the fit of the theoretical models we study. The relative difference between the power spectrum variance from the simulation and the one recovered from the \pin{} mocks is within a few percent while for the bispectrum variance the difference is at the 5\% level (see \OddoEtal{} for further details).

In a fitting problem with $N_p$ free parameters, approximating the $N_b \times N_b$ covariance matrix $\C$ of the data with the sample covariance $\Ce$ measured from a finite number $N_m$ of mock catalogs leads to spuriously enlarged errors for the model parameters.
According to \cite{TaylorJoachimiKitching2013, PercivalEtal2014}, the actual parameter covariance is multiplied by the factor
\be
f=1+\frac{(N_m-N_b-2)(N_b-N_p)}{(N_m-N_b-1)(N_m-N_b-4)}\;.
\ee
In our case, setting $N_m=10\,000$, a maximum of $N_b=233$, and assuming $N_p=10$, gives $f=1.023$. We thus expect that our error estimates for the model parameters are accurate to percent level.

\begin{figure}
    \centering
    \includegraphics[width=0.9\textwidth]{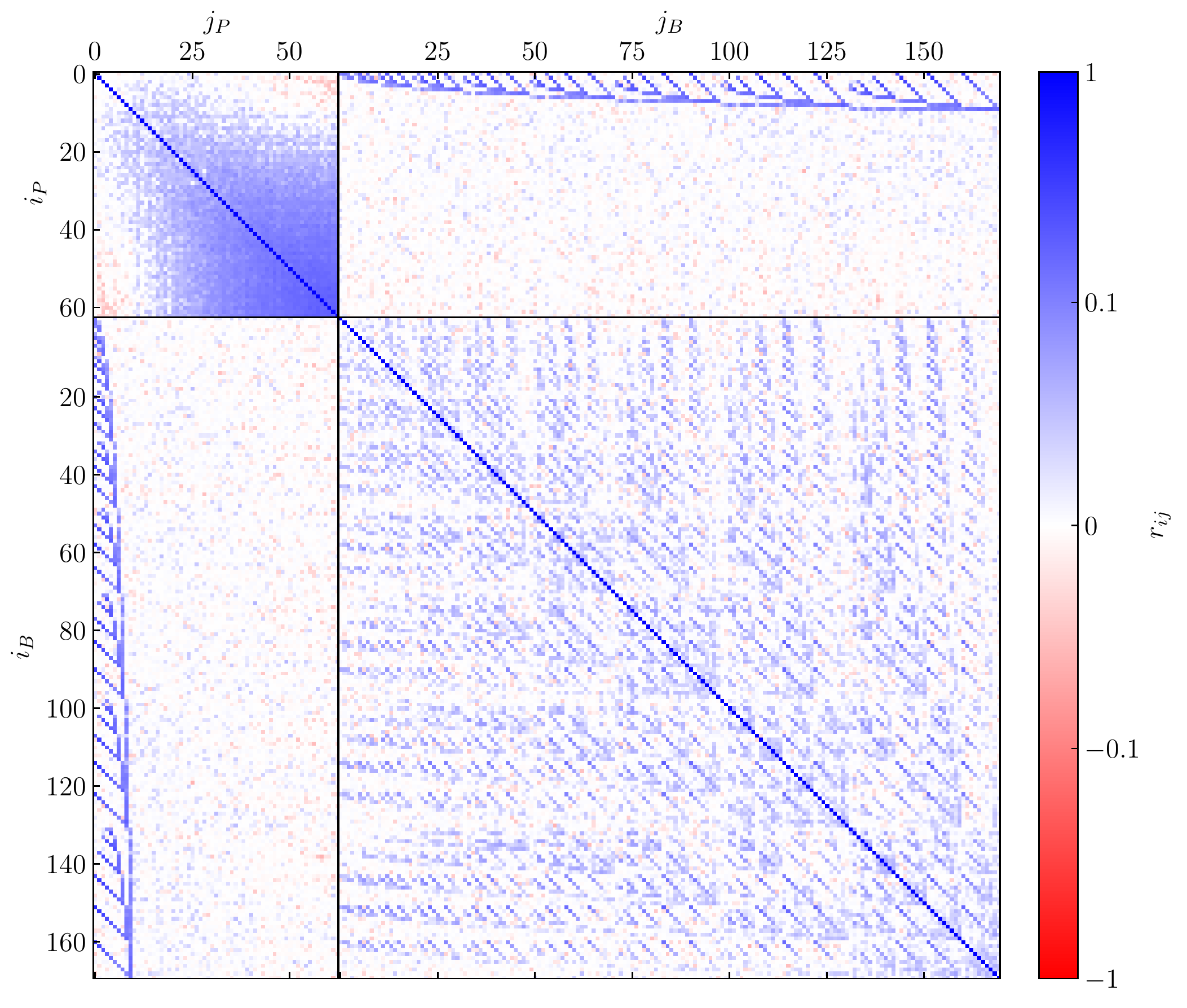}
    \caption{Correlation matrix for the power spectrum and bispectrum measurements of the mock halo catalogs generated with the code \pin{}. The upper-left quadrant shows the power spectrum correlations between the measurements in the 63 Fourier bins from $k_{\rm min,P}\simeq 0.0063 \kMpc$ to $k_{\rm max,P}\simeq 0.53 \kMpc$; the lower-left quadrant shows the bispectrum correlations between the measurements in the 170 triangle bins from $k_{\rm min,B}\simeq 0.0063 \kMpc$ to $k_{\rm max,B}\simeq 0.09 \kMpc$. The other two quadrants show the cross-correlations between the power spectrum and the bispectrum measurements.}
    \label{fig:covariance}
\end{figure}

Figure \ref{fig:covariance} shows the correlation matrix
\be
    r_{ij} = \frac{\widetilde C_{ij}}{\sqrt{\widetilde C_{ii} \widetilde C_{jj}}}
\ee
for the power spectrum and bispectrum measurements estimated from the set of 10\,000 mock halo catalogs. Specifically, the upper left and the lower right quadrants show the correlations of power spectrum and bispectrum respectively, with maximum wavenumbers ${\kmaxP = 0.53 \kMpc}$ for the power spectrum and ${\kmaxB = 0.09 \kMpc}$ for the bispectrum; the other two quadrants show the cross-correlations between power spectrum and bispectrum measurements. Off-diagonal correlations in the power spectrum are of the order of a few percent, and tend to increase up to 15-20 \% at smaller scales due to the relative importance of non-linearities and of the shot-noise in that regime. In the bispectrum, off-diagonal correlations reach 10-20 \%, while cross-correlations between power spectrum and bispectrum can reach 30-40\% for those triangular configurations where, as one can expect, one of the sides coincides with the power spectrum bin. Therefore, neglecting these correlations could in principle lead to inconsistent results in a likelihood analysis. We explore the effects of possible approximations to this covariance matrix on parameters determination in section \ref{ssec:covariance_app}.

%%%%%%%%%%%%%%%%%%%%%%%%%%%%%%%%%%%%%%%%%%%%%%%%
%%%%%%%%%%%%%%%%%%%%%%%%%%%%%%%%%%%%%%%%%%%%%%%%
%%%%%%%%%%%%%%%%%%%%%%%%%%%%%%%%%%%%%%%%%%%%%%%%
\section{Model inference}
\label{sec:model}

%%%%%%%%%%%%%%%%%%%%%%%%%%%%%%%%%%%%%%%%%%%%%%%%
%%%%%%%%%%%%%%%%%%%%%%%%%%%%%%%%%%%%%%%%%%%%%%%%
\subsection{Theoretical model}
\label{ssec:theoreticalmodel}

The theoretical model we consider for the galaxy/halo power spectrum is essentially equivalent to the one employed for the recent analyses of the BOSS data in \cite{IvanovSimonovicZaldarriaga2020, DAmicoEtal2020} and tested in the challenge paper \cite{Nishimichi2012}, albeit limited to real space. This assumes for the matter power spectrum the one-loop expression in Standard Perturbation Theory (SPT) (see, \eg \cite{BernardeauEtal2002}) with the addition of a counterterm contribution accounting for the dynamics of short-scale perturbations as proposed in the Effective Field Theory of the Large Scale Structure (EFTofLSS) \cite{CarrascoHertzbergSenatore2012}. The galaxy power spectrum expression includes galaxy bias one-loop corrections, arising from local and non-local bias operators, and taking into account bias renormalization (see \cite{DesjacquesJeongSchmidt2018} for a recent review). The bispectrum model is limited to the tree-level expression, the leading contribution in SPT. In addition, we account for the damping of the oscillatory features in the power spectrum and the bispectrum following the IR resummation approach of \cite{BaldaufEtal2015B, BlasEtal2016, IvanovSibiryakov2018}. In the following, we write explicitly the expressions we assume for both the power spectrum and the bispectrum. 

We consider the bias expansion for the galaxy overdensity $\delta_g$ given by
\be
    \delta_g = b_1 \delta + \frac{b_2}{2}\delta^2 + b_{\mathcal G_2}\mathcal G_2 + b_{\Gamma_3}\Gamma_3 + b_{\nabla^2 \delta}\nabla^2 \delta + \epsilon + \epsilon_\delta \delta,
    \label{eq:bias}
\ee
where $\delta$ is the matter overdensity, while $\mathcal G_2$ and $\Gamma_3$ are the relevant non-local operators, up to third order, that can be written as a function of the gravitational and velocity potentials $\Phi$ and $ \Phi_v$ as
\bea
    \mathcal G_2 &\equiv& \left[ (\partial_i \partial_j  \Phi)^2 - (\nabla^2  \Phi)^2 \right] \\
    \Gamma_3 &\equiv& \G_2(\Phi)-\G_2(\Phi_v).
\eea
The bias expansion in eq.~(\ref{eq:bias}) includes as well the $\nabla^2 \delta$ higher-derivative operator \cite{McDonaldRoy2009} (that could be particularly relevant for massive halos \cite{FujitaEtal2020, NadlerPerkoSenatore2018}), while $\epsilon$ and $\epsilon_\delta \delta$ are stochastic contributions to the galaxy density field \cite{DekelLahav1999, Matsubara1999, TaruyaSoda1999}. We are not considering those third order operators that only provide scale-independent corrections contributing to the renormalization of linear bias \cite{McDonald2006, AssassiEtal2014, EggemeierScoccimarroSmith2019}. 

The expression for the one-loop galaxy power spectrum in real space can be written as the sum of the SPT model plus contributions due to the higher-derivative bias corrections, the counterterm of the matter power spectrum, and stochasticity
\bea
    P_{gg}(k) &=& P_{\rm SPT}(k) + P_{\rm h.d.}(k) + P_{\rm ct}(k) + P_{\rm stoch}(k).
    \label{eq:Pkterms}
\eea
The SPT model is explicitly given by 
\bea
    P_{\rm SPT}(k)(k) &=& b_1^2\,\Pl(k) + 2\,\int \de^3 \qv\, \left[ K_2(\qv, \kv-\qv) \right]^2\Pl(q) \Pl(\vert \kv - \qv \vert) + \nn\\
    &+& 6\,b_1\,\Pl(k) \int \de^3 \qv\, K_3(\kv, \qv, -\qv) \Pl(q)\,.
\eea
with $\Pl(k)$ being the linear matter power spectrum, and the $K_n$ kernels defined (in analogy to the redshift-space kernels $Z_n$ adopted, e.g., in \cite{BernardeauEtal2002}) as
\bea
    K_2(\kv_1, \kv_2) &=& b_1 F_2(\kv_1, \kv_2) + \frac{b_2}{2} + b_{\mathcal G_2} S(\kv_1, \kv_2) \\
    K_3(\kv_1, \kv_2, \kv_3) &=& b_1 F_3(\kv_1, \kv_2, \kv_3) + \frac{b_2}{3} \left[F_2(\kv_1, \kv_2) + {\rm cyc.} \right] + \nn\\
    &+& \frac{2}{3} b_{\mathcal G_2} \left[S(\kv_1, \kv_{23})F_2(\kv_2, \kv_3) + {\rm cyc.}\right]- \frac{4}{21}b_{\Gamma_3} \left[S(\kv_1, \kv_{23})S(\kv_2, \kv_3) +{\rm cyc.}\right],
\eea
where 
\bea
    F_2(\kv_1, \kv_2) &=& \frac{5}{7} + \frac{1}{2}\left( \frac{k_1}{k_2}+\frac{k_2}{k_1} \right)\hat{\kv}_1\cdot\hat{\kv}_2 + \frac{2}{7}\left( \hat{\kv}_1\cdot\hat{\kv}_2 \right)^2 \,,
\eea
is the usual second-order kernel of the matter expansion in the Eistein-de Sitter approximation, while
\bea
    S(\kv_1, \kv_2) &=& \left( \hat{\kv}_1\cdot\hat{\kv}_2 \right)^2-1,
\eea
provides the tidal term at second order. We refer the reader to, \eg, \cite{GoroffEtal1986} for an explicit expression of the third order kernel $F_3(\kv_1, \kv_2, \kv_3)$ of the matter density SPT solution. 
The higher-derivative bias corrections lead to the galaxy power spectrum contribution 
\be
P_{\rm h.d.}(k) = - 2 b_1 b_{\nabla^2\delta}k^2 \Pl(k)\,
\ee
while the EFT counterterm leads to
\be
P_{\rm ct}(k) = - 2 b_1^2 c_s^2 k^2 \Pl(k)\,,
\ee 
with $c_s^2$ representing the effective sound speed of the matter fluid. Finally, we write the stochastic contribution as
\be
P_{\rm stoch}(k) = \left( 1 + \alpha_P + \epsilon_{k^2}k^2 \right)\bar n^{-1}\,
\ee
where the two free parameters $\alpha_P$ and $\epsilon_{k^2}$ describe, respectively, constant and scale-dependent corrections to the Poisson shot-noise term $\bar n^{-1}$. The integrals include loop-corrections to the matter power spectrum $\Delta P_{\rm 1-loop}^m(k)$, as well as other contributions coming from the galaxy bias expansion. 

We can expand equation~(\ref{eq:Pkterms}) to obtain
\bea
    P_{gg}(k) &=& b_1^2 P_{mm}(k) + b_1 b_2 P_{b_1 b_2}(k) + b_1 b_{\mathcal G_2} P_{b_1 b_{\mathcal G_2}}(k) + b_1 b_{\Gamma_3} P_{b_1 b_{\Gamma_3}}(k) + b_2^2 P_{b_2 b_2}(k) + \nn \\
    &+& b_2 b_{\mathcal G_2} P_{b_2 b_{\mathcal G_2}}(k) + b_{\mathcal G_2}^2 P_{b_{\mathcal G_2} b_{\mathcal G_2}}(k) - 2 b_1 b_{\nabla^2 \delta} k^2 \Pl(k) + \left( 1 + \alpha_P + \epsilon_{k^2}k^2 \right)\bar n^{-1},
    \label{eq:Pgg}
\eea
where 
\be
P_{mm}(k) = \Pl(k) + \Delta P_{\rm 1-loop}^m(k) - 2 c_s^2 k^2 \Pl(k)
\label{eq:Pmm}
\ee
is the one-loop model for the matter power spectrum,
and where we have introduced the individual contributions
\begingroup
\allowdisplaybreaks
\bea
\Delta P_{\rm 1-loop}^m(k) &=& 2\, \int \de^3 \qv\, \left[ F_2(\qv, \kv-\qv) \right]^2\Pl(q) \Pl(\vert \kv - \qv \vert) + \nn\\
&\hspace{0.5em}&\hspace{9em}+6\,\Pl(k) \int \de^3 \qv\, F_3(\kv, \qv, -\qv) \Pl(q)\,, \label{eq:Pb1b1}\\
P_{b_1 b_2}(k)&=& 2\int \de^3 \qv\, F_2(\qv, \kv- \qv)\Pl(q)\Pl(|\kv-\qv|) \label{eq:Pb1b2}\,,\\
P_{b_1 b_{{\mathcal G}_2}}(k)&=& 4\int \de^3 \qv\, F_2(\qv, \kv- \qv)S(\qv, \kv- \qv)\Pl(q)\Pl(|\kv- \qv|) + \nn\\
&\hspace{0.5em}&\hspace{9em}+8 \Pl(k) \int \de^3  \qv\, F_2 (\kv,-\qv) S(\qv, \kv - \qv) \Pl(q) \label{eq:Pb1bG2}\,,\\
P_{b_1 b_{\Gamma_3}}(k)&=&  - \frac{16}{7} \Pl(k)\int \de^3 \qv\, S(\qv, \kv-\qv)S(\kv, \qv)\Pl(q) \label{eq:Pb1bG3}\,,\\
P_{b_2 b_2}(k)&=& \frac{1}{2}\int \de^3\qv\,  \Pl(q)\Pl(|\kv-\qv|) - \frac{1}{2}\int \Pl^2(q) \de^3 q \label{eq:Pb2b2}\,,\\
P_{b_2 b_{{\mathcal G}_2}}(k)&=& 2\int \de^3 \qv\, S(\qv, \kv- \qv)\Pl(q)\Pl(|\kv- \qv|) \label{eq:Pb2bG2}\,,\\
P_{b_{{\mathcal G}_2}b_{{\mathcal G}_2}}(k)&=& 2\int \de^3 \qv\, S^2(\qv, \kv-\qv)\Pl(q)\Pl(|\kv-\qv|)\label{eq:PbG2bG2}\,.
\eea
\endgroup
The constant subtracted to $P_{b_2 b_2}(k)$ in equation (\ref{eq:Pb2b2}) ensures that all loop-corrections converge to zero in the large-scale limit, and allows for the renormalization of the constant shot-noise parameter $\alpha_P$. Notice that, whenever this model is used without any additional information on its parameters, the EFT counterterm and the higher-derivative contribution are perfectly degenerate. When this is the case, we can define the combination 
\be
\tilde c_0 = b_1^2 c_s^2 + b_1 b_{\nabla^2 \delta}\,,
\label{eq:c0}
\ee
reducing the dimensionality of the parameter space.

We account for the smoothing of the acoustic features due to the bulk flow by implementing the IR resummation \cite{BaldaufEtal2015B, BlasEtal2016} in the power spectrum model. The starting point to this is to split the linear power spectrum into a smooth, no-wiggle part $\Pnw(k)$, capturing the broadband shape of the power spectrum, and a wiggly part $\Pw(k)$, describing the baryon acoustic oscillations,
\be
    \PL(k) = \Pnw(k)+\Pw(k)\,.
\ee
We obtain this split by applying the 1D Gaussian filter method described in the appendix of \cite{VlahEtal2016}. Following \cite{BaldaufEtal2015B}, the wiggly part is then suppressed by a damping factor, $e^{-k^2 \Sigma^2}$, with
\be
\Sigma^2 = \frac{1}{6 \pi^2}\int _0 ^{k_{\rm S}} \de q\, \Pnw(q)\left[ 1 - j_0\left( \frac{q}{k_{\rm osc}} \right) + 2 j_2\left( \frac{q}{k_{\rm osc}} \right) \right],
\ee
where the $j_n(x)$ are the spherical Bessel functions, $k_{\rm osc}$ is the BAO wavenumber, with $1/k_{\rm osc} \simeq 103 \Mpc$ for the \mine{} cosmology, and $k_{\rm S}= 0.2\, h\,{\rm Mpc}^{-1}$ is an arbitrary cut-off scale that separates the short and long modes; note that $\Sigma^2$ is only weakly dependent on the choice of $k_{\rm S}$ (see \cite{BaldaufEtal2015B}). The leading order galaxy power spectrum in real space reads then
\be
    \PLO(k) = b_1^2 \left[ \Pnw(k) + e^{-k^2 \Sigma^2}\Pw(k) \right].
    \label{eq:PLO}
\ee
Finally, all loop corrections are computed replacing $\Pl(k)$ with the leading-order power spectrum for the matter perturbations (obtained by setting $b_1 = 1$ in eq. \ref{eq:PLO}) in eqs. (\ref{eq:Pb1b1})-(\ref{eq:PbG2bG2}). These are then multiplied by the respective bias parameters, and finally summed to obtain the IR-resummed loop correction $\Delta P_{\rm IR}^{\rm 1-loop}$. This leads to the next-to-leading order galaxy power spectrum
\be
    \PNLO(k) = b_1^2 \left[ \Pnw(k) + e^{-k^2 \Sigma^2}\left( 1+ k^2 \Sigma^2 \right)\Pw(k) \right]+\Delta P_{\rm IR}^{\rm 1-loop}(k).
\ee

The tree-level galaxy bispectrum in real space can be written as
\bea
    B_{ggg}(\kv_1, \kv_2, \kv_3) &=&   B_{\rm SPT}(\kv_1, \kv_2, \kv_3) + B_{\rm h.d.}(\kv_1, \kv_2, \kv_3) + B_{\rm stoch}(k_1, k_2, k_3) \,,
    \label{eq:Bggg}
\eea
where
\bea
    B_{\rm SPT}(\kv_1, \kv_2, \kv_3) &=& 2\,b_1^2\,K_2(\kv_1, \kv_2)\Pl(k_1)\Pl(k_2) + {\rm cyc.} \nn \\
    &=& 2 b_1^2\left[ b_1 F_2(\kv_1, \kv_2) + \frac{b_2}{2} + b_{\mathcal G_2}S(\kv_1, \kv_2) \right]\Pl(k_1)\Pl(k_2) + {\rm cyc.} \,,
    \label{eq:Bspt}
\eea
while the contributions coming from the higher-derivative operator in the bias expansion (limited to the correction to linear bias) are given by
\bea
    B_{\rm h.d.}(\kv_1, \kv_2, \kv_3) = &-&2 b_1^2 b_{\nabla^2}(k_1^2+k_2^2+k_3^3) F_2(\kv_1, \kv_2)\PL(k_1)\PL(k_2) + {\rm cyc.}\nn \\
    &-&b_1 b_{\nabla^2}(k_1^2+k_2^2)\left[ b_2 + 2 b_{\mathcal G_2} S(\kv_1, \kv_2) \right]\PL(k_1)\PL(k_2) + {\rm cyc.}\,.
    \label{eq:Bhd}
\eea
Although these contributions are often included only in one-loop corrections \cite{EggemeierScoccimarroSmith2019}, we still decide to include them because, since they depend on the same $b_{\nabla^2\delta}$ appearing also in the power spectrum model, they can potentially break the degeneracy between the higher-derivative bias and the effective sound speed in the EFT counterterm. Other higher-derivative contributions could be included in the bispectrum model due to further operators (see e.g. \cite{EggemeierEtal2021}). However, as shown in \OddoEtal{}, the bispectrum model at tree-level is able to describe our measurements in terms of the parameters $b_1$, $b_2$, and $\bG$ at the scales we consider. Therefore, while including for the moment the contributions in eq. \ref{eq:Bhd}, we neglect possible contributions from these extra higher-derivative operators.

Finally, the stochastic contribution to the galaxy bispectrum is given by
\be
    B_{\rm stoch}(k_1, k_2, k_3) = \frac{1+\alpha_1}{\bar n}\,b_1^2\,\left[ \PL(k_1) + \PL(k_2) + \PL(k_3) \right] + \frac{1+\alpha_2}{\bar n^2},
\ee
with $\alpha_1$ and $\alpha_2$ representing corrections to the Poisson prediction.

As with the power spectrum, we perform here the IR resummation by replacing any instance of the linear matter power spectrum $\Pl(k)$ with its IR-resummed counterpart, the leading order power spectrum $\PLO(k)$. Notice that, with respect to e.g. \cite{IvanovSibiryakov2018}, for simplicity we do not subtract the contribution proportional to ${\rm e}^{-2k^2 \Sigma^2}P_w^2$, since it is negligible at the scales we consider.

%%%%%%%%%%%%%%%%%%%%%%%%%%%%%%%%%%%%%%%%%%%%%%%%%%%%%%
\subsection{Fourier-space grid effects}
\label{ssec:binning}

While the theoretical models for matter or halo correlators are functions of Fourier wavevectors defined over an infinite volume, measurements from N-body simulations in boxes with periodic boundary conditions are not. Therefore, care must be taken when comparing the two, especially in our case where statistical uncertainties are significantly small due to the large volume available.

The most consistent approach requires to average exactly the theoretical model over each Fourier bin. In the power spectrum case this amounts to compute
\be
\label{e:PbinExact}
    P_{gg}^{\rm bin}(k) = \frac{1}{N_P(k)}\sum_{\qv \in k}P_{gg}^{\rm th}(q),
\ee
where in practice we replaced the term $\vert \delta_\qv \vert^2 /L^3$ with its expected theoretical mean $P_{gg}(q)$ in the expression for the power spectrum estimator (\ref{eq:PowerEstimator}). In this way the theory is evaluated on the wavenumbers available on the discrete grid characterising the simulation we want to compare with. Similarly for the bispectrum model, we replace $\delta_{\qv_1} \delta_{\qv_2} \delta_{\qv_3}/L^3$ with $B_{ggg}(\qv_1, \qv_2, \qv_3)$ to obtain
\be
\label{e:BbinExact}
    B_{ggg}^{\rm bin}(k_1, k_2, k_3) = \frac{1}{N_B(k_1, k_2, k_3)}\sum_{\qv_1 \in k_1}\sum_{\qv_2 \in k_2}\sum_{\qv_3 \in k_3}\delta_K(\qv_{123})B_{ggg}^ {\rm th}(\qv_1, \qv_2, \qv_3).
\ee

Clearly this approach is numerically demanding, particularly in a likelihood analysis that requires this evaluation at each step of the Markov chain. One common alternative is to evaluate the theoretical model at ``effective'' values of the Fourier wavenumbers, often computed as averages over the bin, both for the power spectrum and the bispectrum. For the power spectrum this definition is unambiguous and unique for any bin of center $k$,
\be
    k_{\rm eff}(k) = \frac{1}{N_P(k)}\sum_{\qv \in k} \vert \qv \vert,
    \label{eq:keffP}
\ee
and allows for a fast evaluation of the theoretical model as
\be
    \label{eq:Peff}
    P_{gg}^{\rm eff}(k) = P_{gg}^{\rm th}(k_{\rm eff}).
\ee
However, for the bispectrum the definition for average Fourier wavenumbers is not unique. Among a couple of possible choices tested in \citetalias{OddoEtal2020}, the one performing best is the one defined on sorted Fourier wavevectors and defined as follows
\bea
    k_{\rm eff,l}(k_1, k_2, k_3) &=&\frac{1}{N_B(k_1, k_2, k_3)}\sum_{\qv_1 \in k_1} \sum_{\qv_2 \in k_2} \sum_{\qv_3 \in k_3} \delta_K(\qv_{123}) \,{\rm max}(q_1, q_2, q_3) \nonumber\\
    k_{\rm eff,m}(k_1, k_2, k_3) &=& \frac{1}{N_B(k_1, k_2, k_3)}\sum_{\qv_1 \in k_1} \sum_{\qv_2 \in k_2} \sum_{\qv_3 \in k_3} \delta_K(\qv_{123}) \,{\rm med}(q_1, q_2, q_3) \nonumber\\
    k_{\rm eff,s}(k_1, k_2, k_3) &=& \frac{1}{N_B(k_1, k_2, k_3)}\sum_{\qv_1 \in k_1} \sum_{\qv_2 \in k_2} \sum_{\qv_3 \in k_3} \delta_K(\qv_{123}) \,{\rm min}(q_1, q_2, q_3).
    \label{eq:keffB}
\eea
In real space, the effective wavenumbers provide a fast evaluation of the theoretical prediction as
\be
\label{e:BbinEff}
    B_{ggg}^{\rm eff}(k_1, k_2, k_3) = B_{ggg}^ {\rm th}\left(k_{\rm eff,l}, k_{\rm eff,m}, k_{\rm eff,s}\right).
\ee

We also consider, as an additional approach, an extension to the effective wavenumbers prescription based on a Taylor expansion of the theoretical model. For the power spectrum we can write, for instance,
\be
    P_{\rm bin}(k) = \frac{1}{N_P(k)}\sum_{\qv \in k} \sum_{n = 0}^\infty \frac{1}{n!}P^{(n)}(\keff)(q-\keff)^n \simeq P(\keff) + \frac{1}{2}P^{\prime \prime}(\keff) \mu_2(k) \equiv P_{\rm exp}(k)\
\ee
where the Taylor series has been truncated to include up to the second-order term and where
\be
    \mu_2(k) \equiv  \frac{1}{N_P(k)}\sum_{\qv \in k}(q-\keff)^2\,.
\ee
We refer to this approach as an ``expansion'' to the effective approach. Details of its implementation to the bispectrum tree-level predictions can be found in Appendix \ref{app:binning}.

Differences between these approximations and the bin-average of the theoretical model are typically larger for small values of the wavenumbers, and in the bispectrum case, particularly pronounced for squeezed triangular configurations. In general, a larger bin width also leads to a worse agreement with the case involving the full bin-average.

An alternative approach to account for grid effects employs the same expressions in eqs. (\ref{e:PbinExact}) and (\ref{e:BbinExact}), but where the sums over Fourier wavevectors are replaced by integrals (see \cite{EggemeierEtal2021} for a fast implementation for the bispectrum). However, we do not consider this approach in our analysis, since this introduces systematic errors comparable to the statistical uncertainties of our datasets on a wide range of scales, as well to the systematics of the effective wavenumbers approach, see Appendix \ref{app:binning}.

%%%%%%%%%%%%%%%%%%%%%%%%%%%%%%%%%%%%%%%%%%%%%%%%%%%%%%
\subsection{Bias relations}
\label{ssec:bias}

In our analysis we consider some relations between the bias parameters in order to reduce the dimensionality of the parameter space. These relations are helpful in the analysis of the real-space power spectrum alone since large degeneracies between $b_{\mathcal G_2}$ and $b_{\Gamma_3}$ make the determination of both bias and cosmological parameters difficult. As shown in section \ref{ssec:cosmores}, this is less of a problem in the joint fit of power spectrum and bispectrum, since the latter provides useful constraints on $b_{\mathcal G_2}$. Still, reducing the dimensionality of the bias parameter space can in general provide tighter constraints on the cosmological parameters, as long as it does not introduce systematic errors relevant for the level of statistical uncertainty that characterises our measurements. 

We test the following relations between bias parameters with joint fits of power spectrum and bispectrum:
\begin{itemize}
    \item the relation $b_2(b_1, b_{\mathcal G_2})$ from \cite{LazeyrasEtal2016}, given by
    \be
    b_2 = 0.412 - 2.143\, b_1 + 0.929\, b_1 ^2 + 0.008\, b_1^3 + \frac{4}{3}\,\bG;
    \label{eq:relationb2}
    \ee
    this is a fitting formula from measurements in separate universe simulations, obtained for values of $b_1$ in the range $(1,10)$; notice that the $4/3\;\bG$ term in the equation accounts for the different definition of the bias expansion adopted in \cite{LazeyrasEtal2016};
    \item the relation $b_{\mathcal G_2}(b_1)$ from \cite{EggemeierEtal2020},
    \be
    \bG = 0.524 - 0.547\, b_1 + 0.046\, b_1^2\,,
    \label{eq:relationbG2}
    \ee
    obtained as a quadratic fit to the excursion set prediction of the tidal bias in \cite{ShethChanScoccimarro2013};
    \item the relation $b_{\Gamma_3}(b_1,b_{\mathcal G_2})$ derived in \cite{EggemeierScoccimarroSmith2019, EggemeierEtal2020} assuming conserved evolution of the galaxies after formation (co-evolution), that in our basis becomes
    \be
    \bgamma = -\frac{1}{6}(b_1-1)-\frac{3}{2}\bG\,.
    \label{eq:relationbG3}
    \ee
\end{itemize}

While this is not a comprehensive list of all possible relations proposed in the literature, they represent a starting point of possible relations to explore. We do not consider the local-Lagrangian relation between $b_{\mathcal G_2}$ and $b_1$ \cite{ChanScoccimarroSheth2012, BaldaufEtal2012} since different studies have shown its limits \cite{LazeyrasSchmidt2018, AbidiBaldauf2018}; moreover, in the bispectrum-only analysis of \citetalias{OddoEtal2020}, it is shown to lead to systematic errors in the constraints on the bias parameters, at least when the full simulation volume is considered.

All these bias relations have been derived and studied in the context of distributions of dark matter halos. While their validity is expected to extend as well, to some degree, to galaxy measurements (see, e.g. \cite{BarreiraLazeyrasSchmidt2021, ZennaroEtal2021A}), we will leave the quantitative exploration of such topic to future work.

%%%%%%%%%%%%%%%%%%%%%%%%%%%%%%%%%%%%%%%%%%%%%%%%%%%%%%
\subsection{Likelihood function}

As for the analysis in \citetalias{OddoEtal2020}, we fit all power spectrum and bispectrum measurements from the \mine{} simulations together, assuming that they are independent. This means that the total log-likelihood we use to sample the parameter space is given by
\be
\log \mathcal L_{\rm tot}(\boldsymbol \theta) = \sum _\alpha \log \mathcal L_\alpha(\boldsymbol \theta \vert \mathbf{X}_\alpha),
\ee
where the subscript $\alpha$ runs over all realizations, $\boldsymbol \theta$ is the parameters vector, and $\mathbf X_\alpha$ is the dataset of realization $\alpha$. The dataset $\mathbf X_\alpha$ represents either the data vector for the power spectrum, the one for the bispectrum, or the combination of the two. 
For the individual $\log \mathcal L_\alpha$, we use two different likelihood functions, depending on the type of covariance used. When the covariance is chosen as the sample covariance of the measurements from the mock catalogs, we assume the Sellentin \& Heavens likelihood \cite{SellentinHeavens2016} to account for the residual uncertainties in the numerical estimation of the precision matrix due to the finite number of mocks. In the case of a theoretical prediction for the covariance (diagonal assuming Gaussianity), we assume the usual Gaussian likelihood. In both cases, the individual likelihood $\mathcal L_\alpha$ can be written as a function of the chi-square of the model for each single realization $\alpha$. This allows for a fast evaluation of the likelihood when only bias parameters are varied.

%%%%%%%%%%%%%%%%%%%%%%%%%%%%%%%%%%%%%%%%%%%%%%%%%%%%%%
\subsection{Likelihood evaluation}

We perform two types of analyses: in the first, the cosmological parameters are fixed to the values used to run the N-body simulations and we perform tests of the different implementations and approximations of the theoretical model; in the second, we also vary three cosmological parameters in order to assess if possible model systematics can bias the recovered cosmological information.

When the cosmological parameters are fixed, the parameter space is given by the set of 10 parameters $\lbrace b_1, b_2, b_{\mathcal G_2}, b_{\Gamma_3}, c_s^2, b_{\nabla^2 \delta}, \alpha_P, \alpha_1, \alpha_2, \epsilon_{k^2}\rbrace$. We refer to this choice as the ``maximal model'', since we explore the possibility of reducing the number of parameters by setting some of them to zero, or by imposing the relations described in section \ref{ssec:bias}.
The priors on these parameters are assumed to be uniform, and are given in table~\ref{tab:priorsPB}. In addition, we consider varying the power spectrum amplitude parameter $A_s$, the Hubble parameter $h$, and the relative matter density parameter $\omega_m = \Omega_m h^2$. Including these cosmological parameters, and in particular $A_s$, introduces degeneracies that can hamper the estimation of the posterior with Monte Carlo Markov Chains (MCMC).
For this reason, we define the relative amplitude parameter $A \equiv A_s / A_s^{\rm fid}$.
Any $n$-th order operator $\mathcal O$ scales proportionally to $A^{n/2}$. In order to reduce the degeneracies between the bias parameters and the amplitude of the scalar perturbations, we redefine the coefficients $b_{\mathcal O}$ as follows:
\be
\label{e:A_params}
\tilde b_1 = A^{1/2}\,b_1,\quad \tilde b_2 = A \,b_2, \quad \tilde b_{\mathcal G_2} = A\, b_{\mathcal G_2}, \quad \tilde b_{\Gamma_3} = A^{3/2}\,b_{\Gamma_3}\,,
\ee
adopting the same uniform priors. We keep the tilt of the scalar power spectrum $n_s$ and the baryon content $\Omega_b$ fixed, as these parameters are very well constrained by CMB experiments. 

\begin{table}[!t]
    \centering

    \begin{tabular}{ll|c}
        \hline
        \hline
        \multicolumn{2}{l}{Parameter} & Prior (uniform)\\
        \hline
        $b_1$ & & $[0.9,3.5]$\\
        $b_2$ & & $[0,4]$\\
        $\bG$ & & $[-4,4]$\\
        $\bgamma$ & & $[-10,10]$\\
        $\tilde c_0$ &$[\sMpc]$ & $[-100,100]$\\
        $b_{\nabla^2\delta}$ &$[\sMpc]$ & $[-100,100]$\\
        $\epsilon_{k^2}$ &$[\sMpc]$ & $[-10,10]$\\        
        $\alpha_P$ & &$[-1,1]$\\        
        $\alpha_1$ &  &$[-1,1]$\\
        $\alpha_2$ &  &$[-1,1]$\\
        \hline
        $A_s/A_s^{\rm fid}$& & $[0.0004,4.]$\\
        $h$ & &$[0.4,1.]$\\
        $\Omega_m h^2$ & &$[0.07224,0.2224]$\\        
        \hline
        \hline
    \end{tabular}
    
    \caption{Uniform prior intervals of the model parameters.}
    \label{tab:priorsPB}
\end{table}

We evaluate posterior distributions by means of MCMC using the code \texttt{emcee} \cite{ForemanMackeyEtal2013}.
With fixed cosmological parameters, we evaluate the posterior distribution by simulating 100 dependent walkers; moves are performed using the affine invariant ``stretch move'' ensemble method from \cite{GoodmanWeare2010} with parallelization, as described in \cite{ForemanMackeyEtal2013}. We run chains for a number of steps equal to $\min(50\,000,\,100 \tau)$, where $\tau$ is the integrated autocorrelation time. With our setup, we can run chains of this type in approximately 5 minutes on a single core of a laptop. When we include cosmological parameters, we evaluate the posterior distribution by simulating independent chains; moves are performed using a Metropolis-Hastings sampler with steps defined by a Gaussian proposal function, with the parameters covariance determined iteratively running chains a few times. MCMC simulations are run until convergence defined by the Gelman-Rubin diagnostic \cite{GelmanRubin1992}, assuming a precision $\epsilon = 0.05$ and a confidence percentile of $95 \%$. The change in sampling method is due to the longer running times when including cosmological parameters. At each step, we call the Boltzmann solver \texttt{CAMB} \cite{LewisChallinorLasenby2000} to compute the linear power spectrum, we compute loop corrections to the power spectrum using a custom implementation based on the \texttt{FAST-PT} code \cite{McEwenEtal2016}, and perform our IR-resummation routine. Grid effects for both power spectrum and bispectrum are accounted for by adopting the approximated approach outlined in appendix \ref{app:binning}. This allows us to have a likelihood evaluation (and thus one MCMC step) in $\sim 1.5 \, {\rm s}$, and therefore to reach convergence in a relatively short time, of the order of 10 hours (running each independent chain on a separate core at the same time).
Marginalized one-dimensional and two-dimensional posterior distributions are shown in triangle density plots generated through the code \texttt{GetDist} \cite{Lewis2019}.

%%%%%%%%%%%%%%%%%%%%%%%%%%%%%%%%%%%%%%%%%%%%%%%%%%%%%%
\subsection{Goodness of fit and model selection}
\label{ssec:goodness}

As a way to compare the quality of the fits we perform, we compute the posterior predictive $p$-value ${\rm ppp}$ and the posterior-averaged reduced chi-square $\langle  \chi^2_\nu \rangle_{\rm post}$. For details on the particular choice of these diagnostics, we redirect the reader to \citetalias{OddoEtal2020}; for the purposes of the present work, it suffices to say that we consider a value of ${\rm ppp} \geq 0.95$ to signal a failure of the model in reproducing a good fit to the data. We compare instead $\langle  \chi^2_\nu \rangle_{\rm post}$ to the corresponding 95 percent (upper) confidence limit associated to a number of degrees of freedom equal to the total number of data points fitted: when $\langle  \chi^2_\nu \rangle_{\rm post}$ is greater than this value, the model fails to describe the data.

\begin{figure}[!t]
    \centering
    \includegraphics[width=0.7\textwidth]{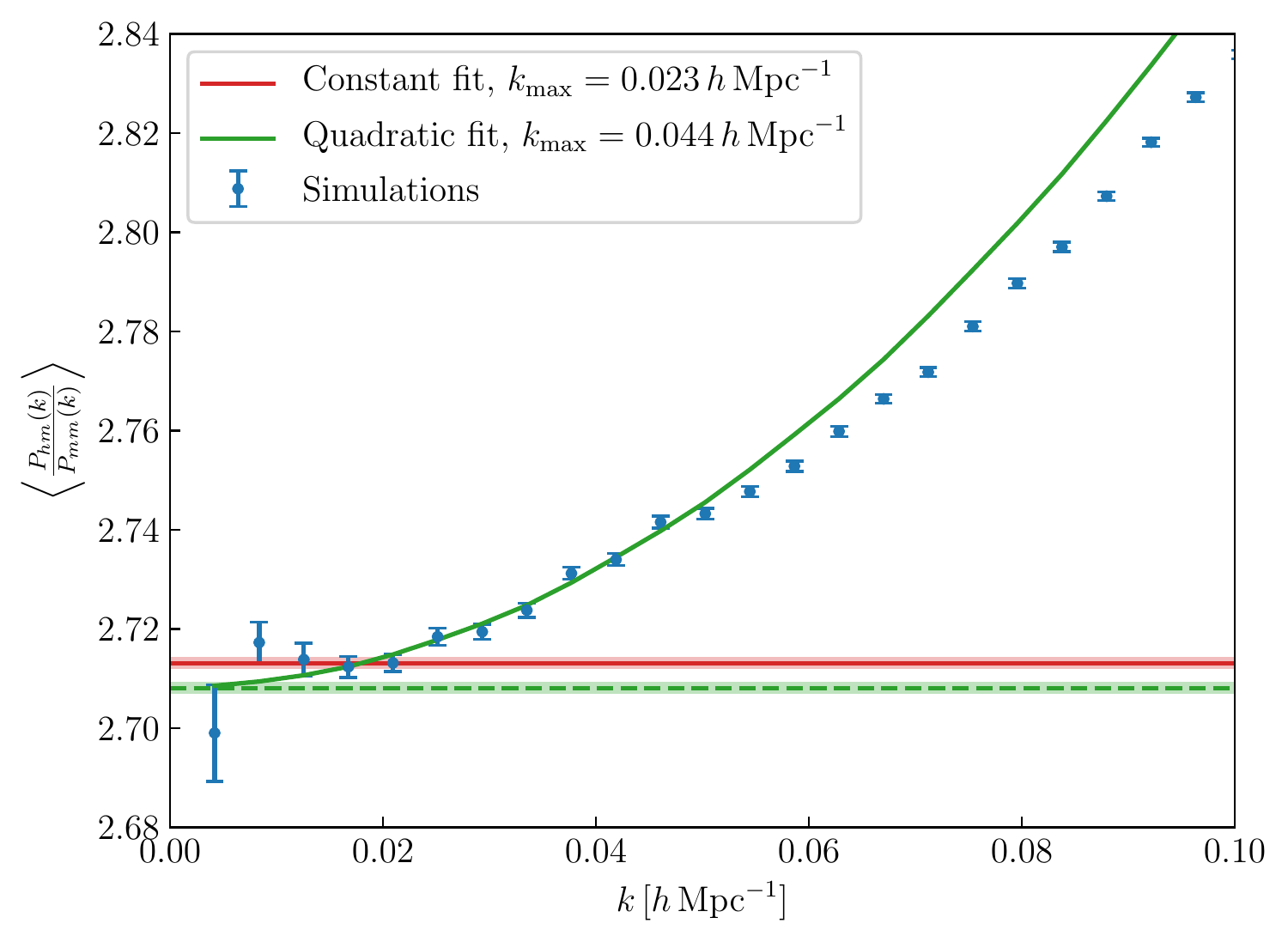}
    \caption{Average of the ratios between the cross halo-matter power spectrum and the matter power spectrum over the full set of \mine{} simulations. The red line is the best fit of the data up to $k_{\rm max} = 0.023\,h\,{\rm Mpc}^{-1}$ with a constant function; the solid green line is the best fit of the data up to $k_{\rm max} = 0.044\,h\,{\rm Mpc}^{-1}$ with a constant plus a $k^2$-dependent term. The green dashed line shows the best-fit value for $b_1$ with this second fit function. Shaded regions show the errors on the best-fit values for $b_1$.}
    \label{fig:b1fit}
\end{figure}

However, our main goal is to extract \textit{unbiased} values of the fitted parameters. For the cosmological sector, the systematic shift can be easily quantified  by comparing the results of the fit to the input values used in the N-body simulations. Conversely, this procedure cannot be followed for the bias parameters as we do not know their values \textit{a priori}. For the sake of understanding whether our analysis leads to biased estimates for the bias sector, we thus attempt to measure $b_1$ from the ratio between the halo-matter cross-power spectrum $P_{hm}(k)$ and the matter auto-power spectrum $P_{mm}(k)$.
In order to obtain an estimate of $b_1$ which is independent of our likelihood pipeline, we fit the large-scale behaviour of the ratio $P_{hm}(k)/P_{mm}(k)$ assuming for the cross-power spectrum the model
\be
P_{hm}(k) = ( b_1 + c k^2 ) P_{mm}(k)
\label{eq:fitcross}
\ee
where $c$ is a constant and the $k^2$-correction is a way to partially account for non-linearities detectable even at the largest scales, see figure \ref{fig:b1fit}. In what follows, we refer to the estimate of $b_1$ derived assuming this model as $b_1^\times$. Using chi-square minimization, we find $b_1 ^\times = 2.7081 \pm 0.0012$, which we use as a reference value\footnote{It is worth stressing that a different fit for $b_1^\times$ was used in \OddoEtal{}, where we fit a constant linear bias coefficient to the ratio $P_{hm}(k)/P_{mm}(k)$ up to $\kmax=0.023\kMpc$. This leads to a best-fit value almost $3\sigma$ away from (and thus inconsistent with) our reference $b_1^\times$. In any case, due to the larger relative uncertainties in \citetalias{OddoEtal2020}, the posteriors of the bispectrum-only analysis are compatible with both values in the same range of validity of the model.} to draw conclusions about the unbiasedness of the posterior distributions extracted from the MCMC runs discussed in section \ref{sec:results}. In practice, we use $b_1^\times$ as if it was the true value for $b_1$.

A note is in order concerning this test. Eq. (\ref{eq:fitcross}) is not equivalent to the full one-loop expression in perturbation theory, given the bias expansion in eq. (\ref{eq:bias}). While the two models share the same large-scale limit, they describe non-linearities in different ways, which could lead to a slightly different value of the linear bias. For this reason, we will not conclude that the results of our main analysis are biased unless they lie more than two standard deviations away from the best-fitting value for $b_1^\times$. Note that our estimate for $b_1^\times$ is only used to measure the bias of our fits and does not enter our fitting procedure as a prior.

An alternative approach in order to test for the consistency of our results with $P_{hm}(k)$ would be to include the cross-power spectrum in the data vector for the likelihood analysis (see, e.g. \cite{EggemeierEtal2020}). However, this would require the full covariance for all correlators, including all cross-covariances, and in our case this is not available.

For the comparison between different models and different assumptions on the bias parameters, we take advantage of the Deviance Information Criterion (DIC) computed from the MCMC simulations as a model-selection statistic. Again, we refer the reader to \citetalias{OddoEtal2020} for a brief introduction to the DIC and a description of our implementation.

%%%%%%%%%%%%%%%%%%%%%%%%%%%%%%%%%%%%%%%%%%%%%%%%
%%%%%%%%%%%%%%%%%%%%%%%%%%%%%%%%%%%%%%%%%%%%%%%%
\section{Results}
\label{sec:results}

We now present the results of our analysis of the halo power spectrum and bispectrum measurements using, unless otherwise stated, the full volume of the combined \mine{} simulations of about 1000$\cGpc$. We stress that, while clearly out of reach for even future surveys, such large volume is still useful to explore and quantify systematic errors from both the model and the methodology.

%%%%%%%%%%%%%%%%%%%%%%%%%%%%%%%%%%%%%%%%%%%%%%%%
\subsection{Selecting the fiducial model}

\begin{figure}
    \centering
    \includegraphics[width=\textwidth]{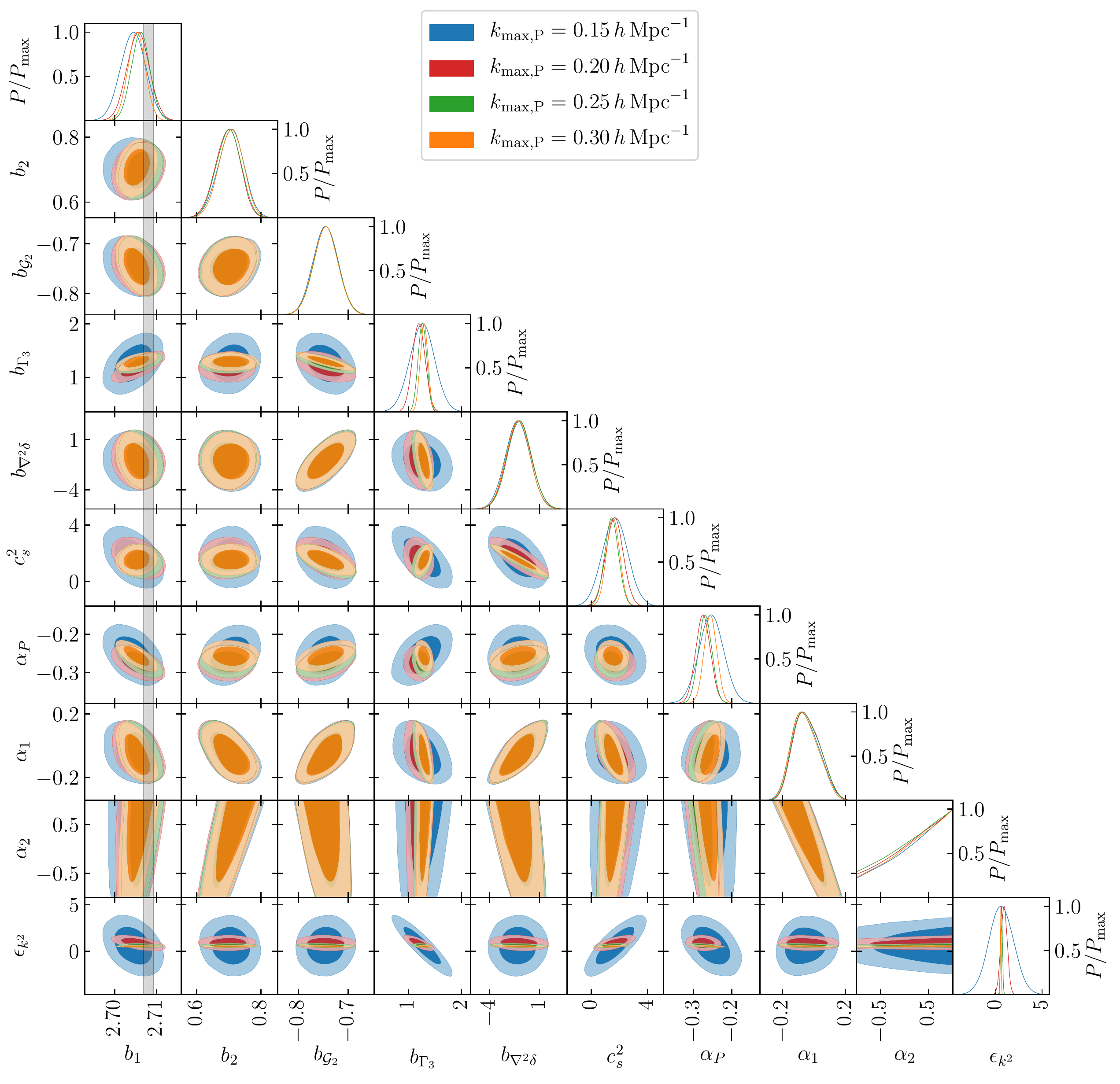}
    \caption{1D and 2D marginalized posteriors for the parameters of the maximal model, obtained through a joint analysis of the halo power spectrum and bispectrum in real space measured from the full set of N-body simulations. The value for $\kmaxB$ is fixed to be $\kmaxB = 0.09 \kMpc$, while the value of $\kmaxP$ is varied; here we show four different values of $\kmaxP = 0.15 \kMpc, 0.20 \kMpc, 0.25 \kMpc, 0.30 \kMpc$ respectively in blue, red, green, and orange. The gray band shows the value of $b_1$ measured from the halo-matter cross-power spectrum, along with its error bar.}
    \label{fig:contourmaximal}
\end{figure}

We first perform a joint analysis of power spectrum and bispectrum adopting the models in equations (\ref{eq:Pgg}) and (\ref{eq:Bggg}) as a function of 10 free parameters: 5 bias coefficients, one EFT counterterm, two stochastic parameters for the power spectrum, and two for the bispectrum. This is what we introduced as the maximal model. Figure \ref{fig:contourmaximal} shows the corresponding 1D and 2D marginalized posteriors with different values of the maximum wavenumber for the power spectrum, $\kmaxP = 0.15$, $0.20$, $0.25$ and $0.30 \kMpc$, while for the bispectrum we consider the fixed maximum wavenumber of $\kmaxB = 0.09 \kMpc$, this being the reach for the tree-level bispectrum model as shown in \OddoEtal. In the rest of this paper, unless otherwise stated, we keep $\kmaxB$ fixed to this value. The recovered value of $b_1$ is consistent, well within the 95\% credibility regions, with the reference value from the cross-power spectrum $b_1^\times$, shown with its own uncertainty by the vertical gray band. We observe that the credible regions for the parameters $\bgamma$, $c_s^2$, $\alpha_P$, and $\epsilon_{k^2}$ shrink as a function of $\kmaxP$. Notice how these parameters do not appear in the bispectrum model, and thus are constrained by the power spectrum alone, so that their constraints improve with larger $\kmaxP$. On the other hand, the constraints on the parameters $b_2$, $\bG$, $b_{\nabla^2\delta}$, $\alpha_1$, and $\alpha_2$ do not improve as a function of $\kmaxP$. The two shot-noise parameters are in fact only present in the bispectrum model (and as such their constraints do not improve with increasing $\kmaxP$, since $\kmaxB$ is kept fixed), while the others appear in the models for both power spectrum and bispectrum, suggesting that they are mostly constrained by the bispectrum.

Just like in the bispectrum-only analysis of \citetalias{OddoEtal2020}, $\alpha_2$ is completely unconstrained inside the prior, and $\alpha_1$ is consistent with zero at $1\sigma$ level. Moreover, while the bispectrum appears to be able to partially break the degeneracy between $\bk$ and $c_{s}^2$, $\bk$ is still consistent with zero. Finally, also the $k^2$ correction to the power spectrum stochasticity is consistent with zero within the 95 percent credibility regions, although only for $\kmaxP=0.15\kMpc$. All of the other parameters are either required by the data, or significantly different from zero. 

\begin{figure}
    \centering
    \includegraphics[width=0.9\textwidth]{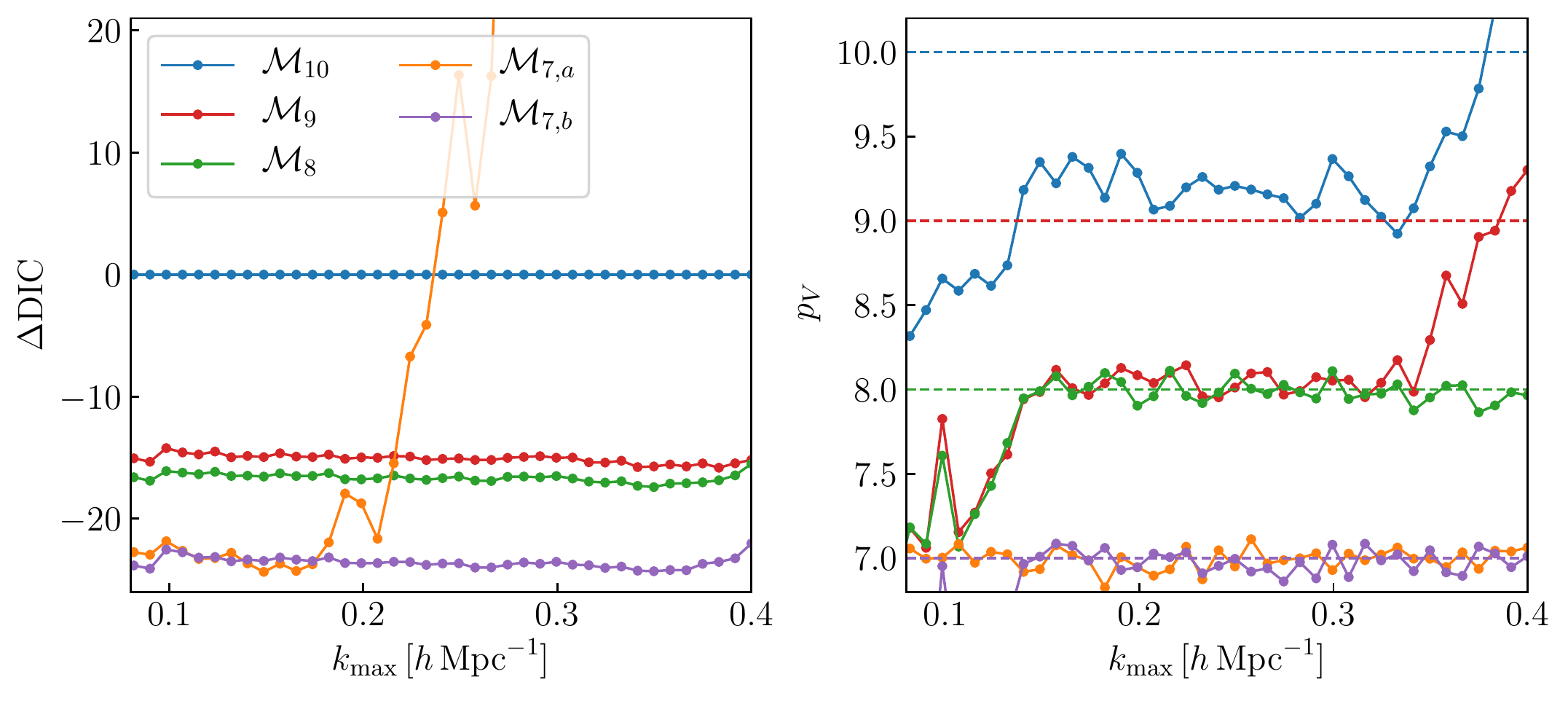}
    \caption{Left panel: difference in DIC of the models considered in the comparison with respect to the maximal model, as a function of $\kmaxP$. Right panel: number of effective parameters $p_V$ constrained by the data for each model, as a function of $\kmaxP$; dashed lines show the actual number of free parameters in the model indicated with the corresponding color (the orange and the purple dashed lines are on top of each other).}
    \label{fig:DICfiducial}
\end{figure}

In order to explore the possibility of a smaller parameter space, we use the DIC to assess the performances of a number of different reductions of the maximal model where a subset of the parameters is set to zero. Figure \ref{fig:DICfiducial} shows the results of the comparison of the following cases:
\begin{itemize}
    \item the maximal model, $\mathcal M_{10}$, with all 10 parameters left free to vary;
    \item the maximal model, $\mathcal M_{9}$, where $\bk$ has been set to zero;
    \item the maximal model, $\mathcal M_{8}$, where both $\bk$ and $\alpha_2$ have been set to zero;
    \item the maximal model, $\mathcal M_{7,a}$, where $\bk$, $\alpha_2$ and $\epsilon_{k^2}$ have been set to zero;
    \item the maximal model, $\mathcal M_{7,b}$, where $\bk$, $\alpha_1$ and $\alpha_2$ have been set to zero.
\end{itemize}
In the left panel, we show the difference in DIC between each model and the maximal model as a function of the $\kmax$ of the power spectrum ($\kmaxB$ being fixed at $0.09\kMpc$). The right panel shows the number of effective parameters $p_V$ constrained by each model, defined as half of the posterior variance of the deviance $D = -2\log \mathcal L_{\rm tot}$ \cite{GelmanEtal2004}, again as a function of $\kmaxP$. 

The large degeneracy between the higher-derivative bias $\bk$ and the EFT counterterm amplitude $c_s^2$ is such that setting $\bk = 0$ allows for a large reduction of the DIC by $\sim 15$ at all values of $\kmax$ considered. For this reason, in all subsequent analysis we set $\bk = 0$, and thus ignore the higher-derivative bias correction in the bispectrum model, while in the power spectrum model it remains as degenerate with the EFT counterterm. In the following, we consider the combination in eq.~(\ref{eq:c0}).
Setting $\alpha_2$ to zero does not improve by much the DIC, however, since $\alpha_2$ is unconstrained and prior dominated, it allows for the number of effective parameters $p_V$ to be consistent with the number of model parameters.

A further reduction to a seven-parameters model can be achieved by setting either $\alpha_1$ or $\epsilon_{k^2}$ to zero, in addition to $\bk = \alpha_2 = 0$. While $\alpha_1 = 0$ decreases the DIC by $\sim 5$ at all values of $\kmaxP$ considered, setting $\epsilon_{k^2} = 0$ leads to a comparable improvement only for $\kmaxP<0.2\kMpc$, but this improvement is rapidly lost for larger Fourier modes. In fact, at small scales the $k^2$ stochastic term turns out to be relevant, perhaps accounting as well for additional corrections beyond the one-loop model we assumed. 

In what follows we consider $\mathcal M_{7,b}$ as the reference bias model, being the one defined by the seven parameters
\be
\label{e:ref_params}
    \boldsymbol\theta_{\rm reference} = \lbrace b_1, b_2, \bG, \bgamma, \tilde c_0, \alpha_P, \epsilon_{k^2}  \rbrace,
\ee
with $\bk$, $\alpha_1$, and $\alpha_2$ set to zero in the bispectrum model. For this case the number of effective parameters shown in the right panel of figure~\ref{fig:DICfiducial} matches the number of free parameters over the entire $\kmaxP$ interval explored.

%%%%%%%%%%%%%%%%%%%%%%%%%%%%%%%%%%%%%%%%%%%%%%%%
\subsection{Analysis with the reference model}
\label{ssec:reference}

\begin{figure}[t]
    \centering
    \includegraphics[width=\textwidth]{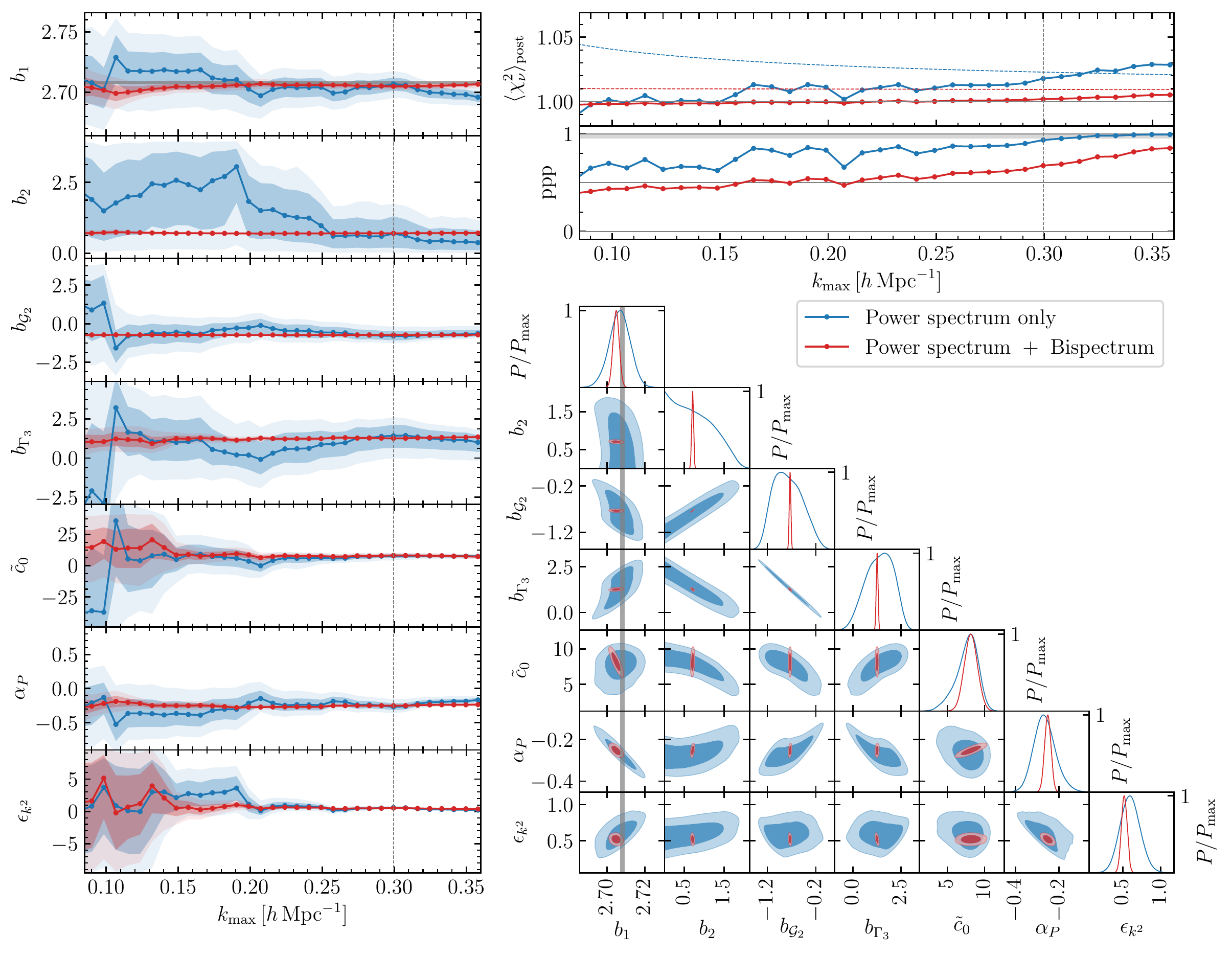}
    \caption{Comparison between the fit of the halo power spectrum and the joint fit of the halo power spectrum and bispectrum measured from the N-body simulations. 
    The left panels show the posterior mean (solid lines), and central 68 and 95 percent ranges (darker and lighter shaded areas respectively) of the model parameters as a function of $\kmax\equiv\kmaxP$. The bispectrum data range is fixed by $\kmaxB=0.09\kMpc$. The vertical dashed line highlights the reference scale of $\kmaxP = 0.30 \kMpc$ for which we display contour plots for the joint posterior density of parameter pairs in the lower-right panel. Here, darker and lighter shaded areas represent the 68 and 95 percent joint credibility regions, respectively. The narrow gray bands indicate the constraints on the linear-bias parameter derived from the halo-matter cross power spectrum. Two goodness-of-fit diagnostics are displayed in the top-right panel as function of $\kmax$: the reduced $\chi^2$ statistic averaged over the posterior (top inset) and the $\rm ppp$ (bottom inset). As a reference, the dashed curves in the top inset indicate the upper one-sided 95 percent confidence limit in a frequentist $\chi^2$ test (note that the number of datapoints included in the fit varies with $\kmax$).}
    \label{fig:referencemodel}
\end{figure}

Figure \ref{fig:referencemodel} shows the results obtained fitting the reference model to the power spectrum data and to the combination of power spectrum and bispectrum data. The left column of panels shows the mean of the posteriors of the model parameters as a function of $\kmaxP$ ($\kmaxB$ is again set to $0.09 \kMpc$).
Darker shaded regions correspond to the central 68 and 95 percent ranges.

The two outcomes are in general agreement, for the most part because of the large uncertainties characterising the posteriors from the power spectrum-only analysis. When fitting the power spectrum alone, we notice that some of the priors turn out to be informative, with $b_2$ being unconstrained from below, and with $\epsilon_{k^2}$ being basically unconstrained at small values of $\kmaxP$, where the $k$-dependent stochastic contribution is expected to be negligible. Moreover, the recovered value of $b_1$ mildly runs as a function of $\kmaxP$, but this feature is not present in the joint analysis. In any case, the posteriors of $b_1$ in the two fits are in general agreement with each other, and they also agree with the value of $b_1^\times$ from the cross halo-matter power spectrum, albeit at $2\sigma$ in the case of the joint-fit. However, considering the method used to fit $b_1^\times$ (see section \ref{ssec:goodness}), we do not deem this deviation to be significant.

The top-right panels show the goodness-of-fit for the power spectrum only (in blue) and for the combination of power spectrum and bispectrum (in red). Both the posterior-averaged reduced chi-square and the ${\rm ppp}$ agree in assessing that the power spectrum model provides a good fit to the data up to $\kmaxP\sim 0.30 \kMpc$, while in the case of the joint fit the model seems to provide a good fit of the data even beyond that.
By direct inspection of the total posterior-averaged chi-squares, compared to the one of a bispectrum-only fit at $\kmaxB = 0.09 \kMpc$, we suspect that this apparent inconsistency is likely due to the large number of triangles in the bispectrum, that reduce the relative weight of the power spectrum in the evaluation of the joint fit.

The bottom-right panels show 1D and 2D marginalized posteriors for the power spectrum-only fit and for the joint fit, with $\kmaxP = 0.30 \kMpc$. A number of results are worth noticing. The addition of the bispectrum tightens the constraints on $b_1$ by a factor of $\sim 3.4$, and is able to break the degeneracies between $b_2$ and $b_{\mathcal G_2}$, and between $b_{\mathcal G_2}$ and $b_{\Gamma_3}$, thus providing a significant improvement in constraining higher-order bias parameters. The constraints on the effective counterterm $\tilde c_0$ are shrunk by almost a factor of two, while the ones on $\alpha_P$ and $\epsilon_{k^2}$ by almost a factor of $3.5$.

\begin{figure}[t]
    \centering
    \includegraphics[width=0.95\textwidth]{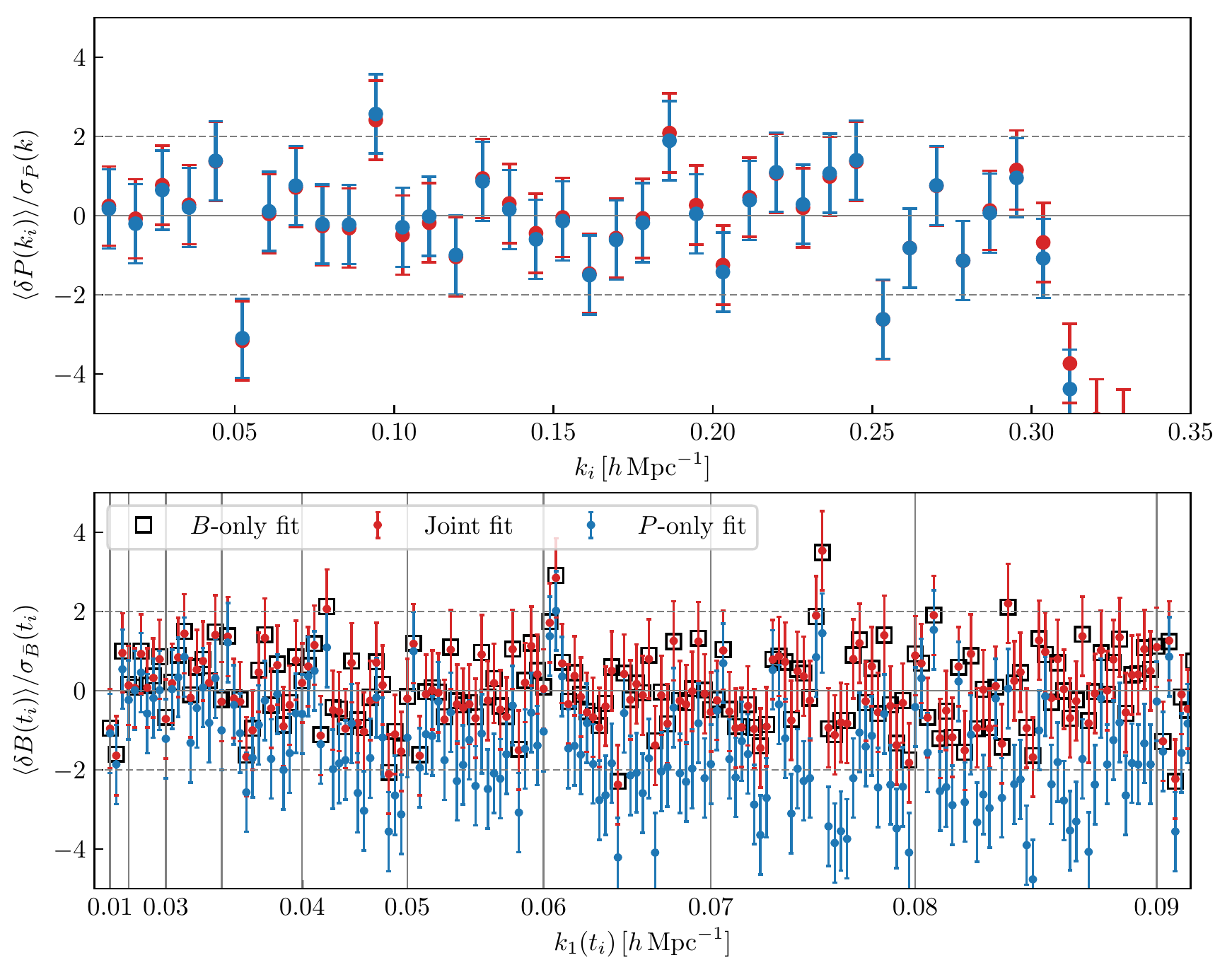}
    \caption{Top: mean residuals, normalized to the standard error on the mean, between the power spectrum measured from the N-body simulations and the posterior-averaged theoretical model from a power spectrum-only fit (blue) and from a joint fit (red), with $\kmaxP = 0.30 \kMpc$ and $\kmaxB = 0.09 \kMpc$. Bottom: same as the top panel, but for the bispectrum measurements; empty squares show the same quantity for the posterior-averaged theoretical model for a bispectrum-only fit with $\kmaxB = 0.09 \kMpc.$ }
    \label{fig:PPC}
\end{figure}

Figure \ref{fig:PPC} shows the comparison between the model from the MCMC fit and the measured data of power spectrum and bispectrum. We compute the posterior-averaged models for both correlators, and then plot the mean residuals with the data, normalized by the standard deviation. The blue markers show points where the model is computed from the posterior of a power spectrum-only fit, while the red markers indicate that the model is computed as the average over the posterior of a joint fit of power spectrum and bispectrum; for the fits in this plot, we choose $\kmaxP = 0.30 \kMpc$ and $\kmaxB = 0.09 \kMpc$. In the plot showing the residuals for the bispectrum, we mark with empty squares the residuals computed with a posterior-averaged model from a bispectrum-only fit. The two cases relative to the power spectrum are consistent up to the $\kmaxP$ of the fit, which is as well close to the maximum Fourier wavenumber up to which the model is expected to work. As shown by the residuals in the lower panel of figure \ref{fig:PPC}, a bispectrum model determined as the posterior-average of a power spectrum-only fit is visibly not able to reproduce the bispectrum data; notice however that the model from the joint fit is largely consistent with the model from a bispectrum-only fit. Since we average the models over the posterior of the MCMC runs, and the posterior from the joint fit is consistent with the posterior from the power spectrum-only fit, we suspect that the observed deviation is due to the additional information provided by the bispectrum on nonlinear bias.

%%%%%%%%%%%%%%%%%%%%%%%%%%%%%%%%%%%%%%%%%%%%%%%%
\subsection{Testing bias relations}

\begin{figure}
    \centering
    \includegraphics[width=\textwidth]{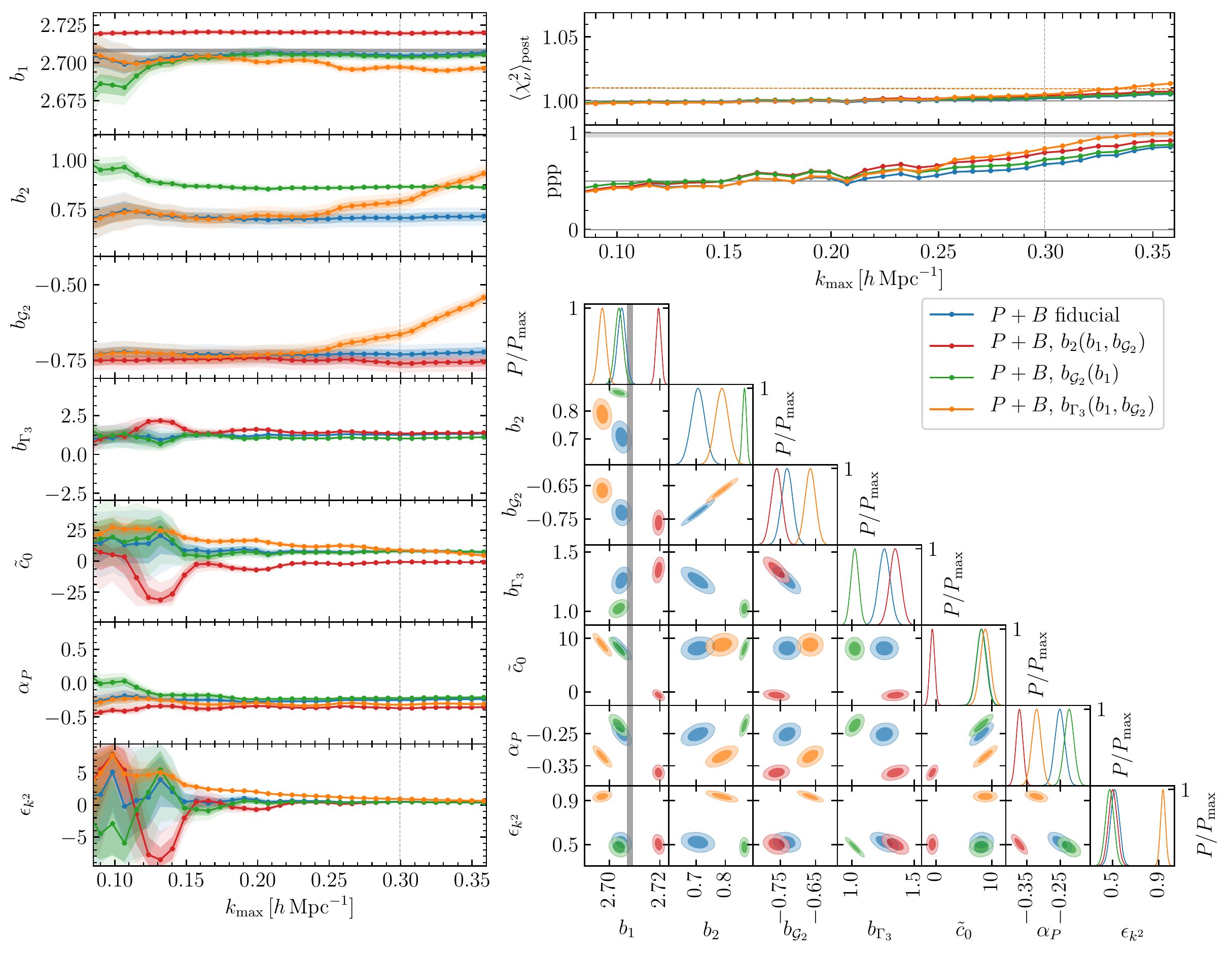}
    \caption{Same as figure \ref{fig:referencemodel}, but comparing the fit performed with the fiducial model to models where different relations between bias parameters are assumed; in blue the reference case of the fiducial model, in red the case where equation (\ref{eq:relationb2}) is assumed, in green the case where we set equation (\ref{eq:relationbG2}), and in orange the case with equation (\ref{eq:relationbG3}).}
    \label{fig:biasrelations}
\end{figure}

We now turn to the performances of the bias relations in eq.s (\ref{eq:relationb2}), (\ref{eq:relationbG2}), and (\ref{eq:relationbG3}). We take our reference model as a starting point and impose each of these relations to reduce to six the number of free parameters. We assume that all relevant halo bias coefficients are physical parameters, consistently describing nonlinear corrections in both the power spectrum and bispectrum, as opposed to simple nuisance parameters. Therefore, any valid physical relation among them should not, in principle, introduce any significant deviation in their recovered values.
The results are shown in figure \ref{fig:biasrelations}. In general, all relations appear to fail, to some extent, in reproducing the values of the parameters obtained with the reference model. We should remember that this test takes advantage of the full simulation volume, well beyond the typical size even of future redshift surveys. It is interesting to notice how the $b_{\Gamma_3}(b_1,b_{\mathcal G_2})$ relation (\ref{eq:relationbG3}) introduces a notable dependence on $\kmaxP$ in the posteriors for parameters like $b_1$ and $b_2$. On the other hand, the $b_{\mathcal G_2}(b_1)$ relation (\ref{eq:relationbG2}) is recovering correctly the expected value of $b_1$, but leading to differences as large as 30\% on parameters as $b_2$. These inconsistencies appear even more significant in the 2D marginalised posteriors in the bottom right inset obtained for $\kmaxP=0.3\kMpc$.

\begin{figure}
    \centering
    \includegraphics[width=\textwidth]{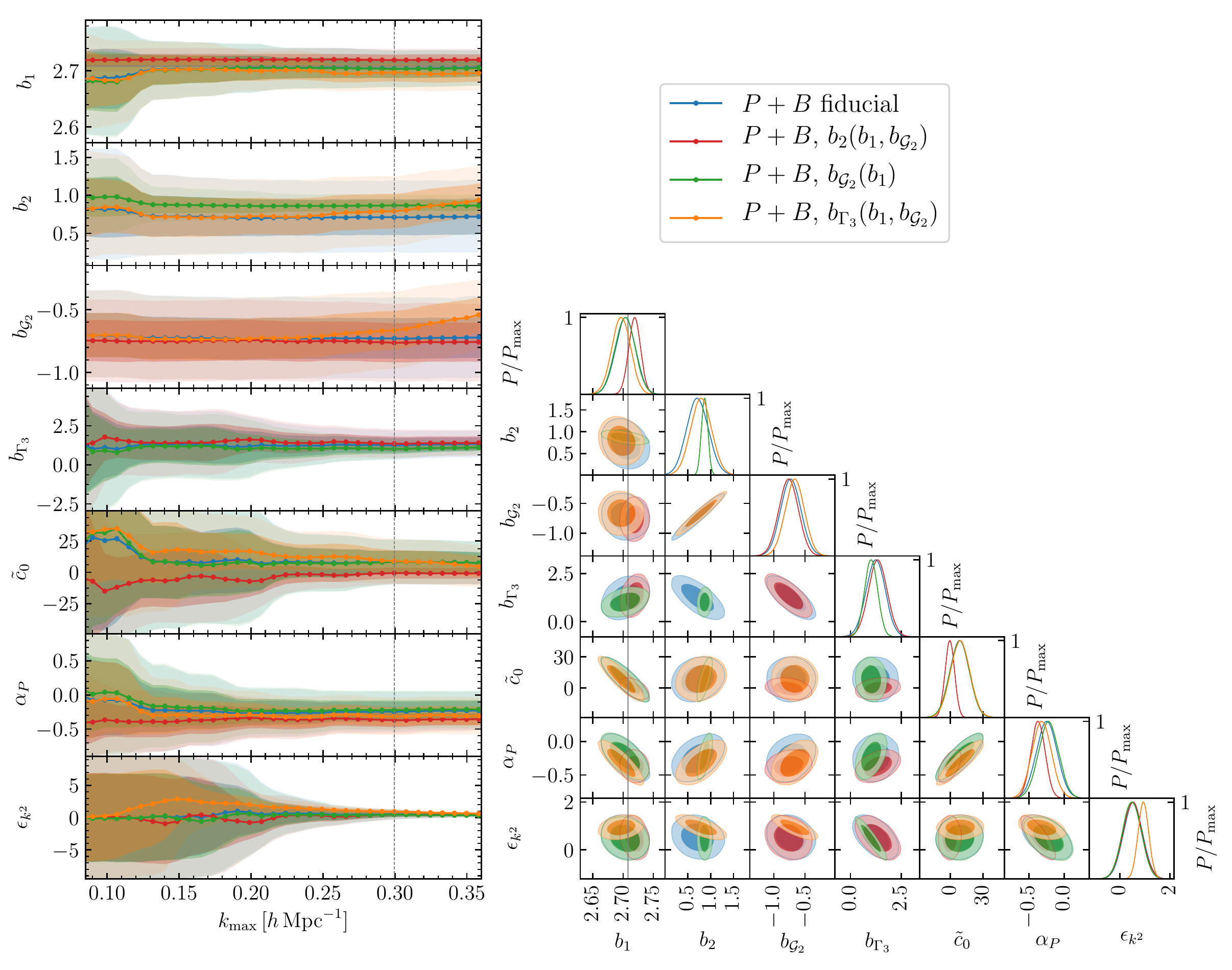}
    \caption{Same as figure \ref{fig:biasrelations}, but for an effective volume of $6 \cGpc$.}
    \label{fig:biasrelationsVeff}
\end{figure}

In order to assess the relevance of the systematic errors induced by the bias relations in a more realistic context, we repeat the same analysis for a smaller effective volume of $V_{\rm eff} = 6 \cGpc$. The effective volume of the full \mine{} dataset is given by (\eg{} \cite{SeoEisenstein2003})
\be
V_{\rm eff}^{\rm N-body}(k) =  \left[ \frac{\bar n P_{hh}(k)}{1+\bar n P_{hh}(k)} \right]^2 V_{\rm N-body},
\ee
where $V_{\rm N-body} = 298 \times (1.5 \Gpc)^3 \simeq 1000 \cGpc$. We then choose the reference $k_r = 0.1 \kMpc$, and then compute the factor $\eta = V_{\rm eff}^{\rm N-body}(k_r)/6 \cGpc$, that we use to rescale the covariance matrix. Finally, we rerun the analysis with the rescaled covariance matrix. The results are shown in figure \ref{fig:biasrelationsVeff}. Notice that, in this case, the goodness-of-fit statistics we have defined cannot be used anymore to determine the range of validity of the model, because of the artificial rescaling of the covariance. For this reason, we simply assume the range of validity to be $\kmaxP = 0.30 \kMpc$. With this smaller effective volume, all bias relations are consistent with the reference analysis and they all provide tighter constraints on one or more parameters, with the $b_2(b_1,\bG)$ fitting function, in particular, leading to the smaller uncertainty on the linear bias $b_1$.

\begin{figure}
    \centering
    \includegraphics[width=0.5\textwidth]{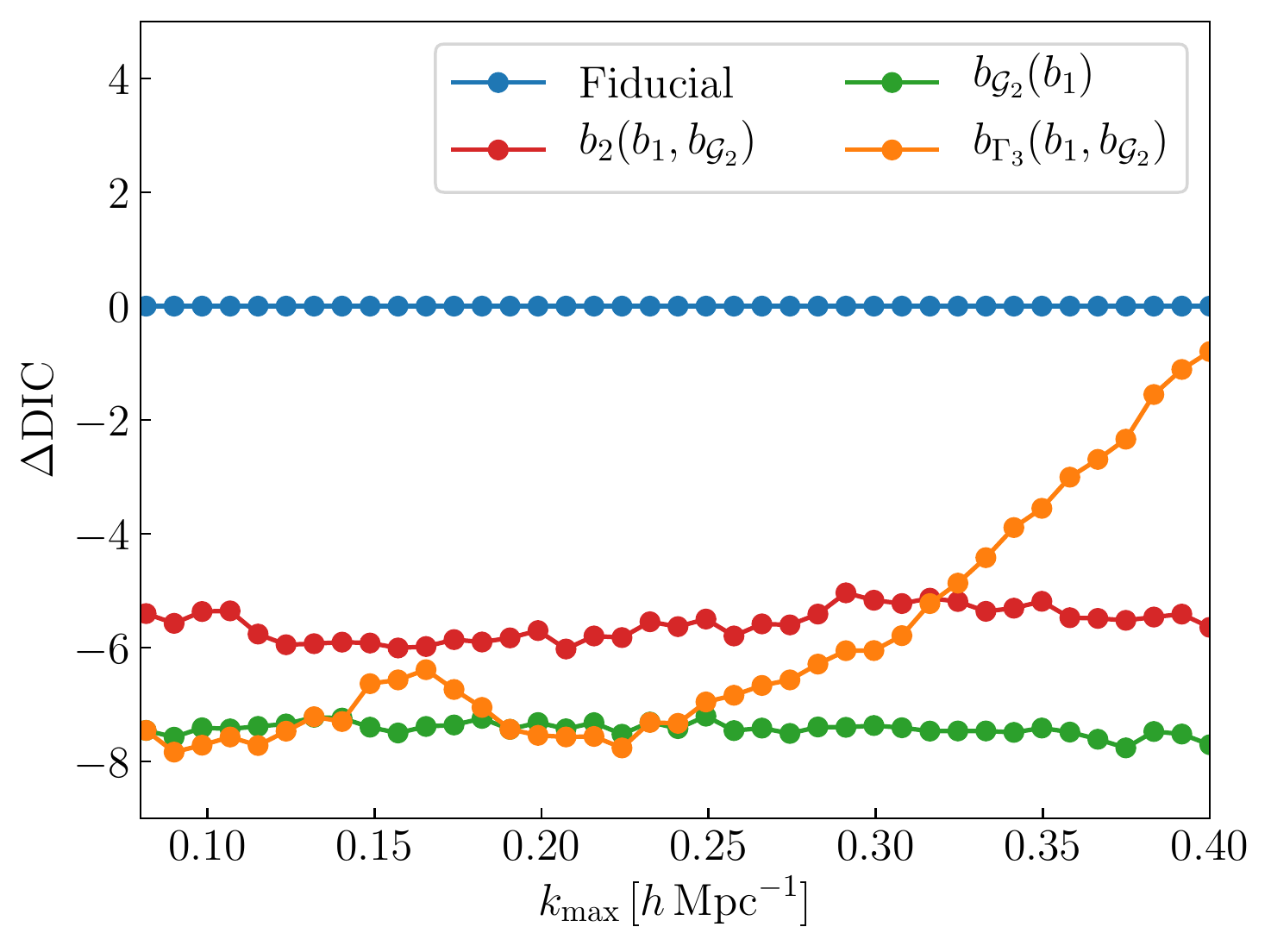}
    \caption{Difference in DIC, with respect to the fiducial case, of the alternative models that assume relations between bias parameters, as a function of $\kmax$, for an effective volume of $6 \cGpc$.}
    \label{fig:DICbias}
\end{figure}

In order to further compare the performance of the bias relations considered, we compute the difference in DIC with respect to the reference analysis, still using the smaller effective volume of $6 \cGpc$, and show them in figure \ref{fig:DICbias}. In the range of validity of the model, all bias relations are favoured with respect to the reference model with the eq.~(\ref{eq:relationbG2}) providing the largest improvement (largest negative difference $\Delta$DIC) over the whole range in $\kmax$. The $b_{\Gamma_3}(b_1,b_{\mathcal G_2})$ relation (\ref{eq:relationbG3}), instead, appears to improve the fit only at the largest scales. 

%%%%%%%%%%%%%%%%%%%%%%%%%%%%%%%%%%%%%%%%%%%%%%%%
\subsection{Effects of binning approximations}
\label{ssec:res_binning}

We now study how different ways to account for Fourier-space grid effects in the theoretical models impact parameter posteriors. Our reference case is the full bin-average of the theoretical predictions of both the power spectrum and the bispectrum, eq.s~(\ref{e:PbinExact}) and (\ref{e:BbinExact}), which we compare to two, more efficient alternatives. We refer to the first as the ``effective wavenumbers'' approach, eq.s ~(\ref{eq:Peff}) and (\ref{e:BbinEff}), where the theoretical predictions are evaluated on the average Fourier wavenumbers, and to the second, based on a Taylor expansion about the effective method approximation, as the ``expansion'' approach, eq.s~(\ref{eq:Pfullexp}) and (\ref{eq:Bfullexp}), both truncated to include up to second-order terms.

\begin{figure}
    \centering
    \includegraphics[width=\textwidth]{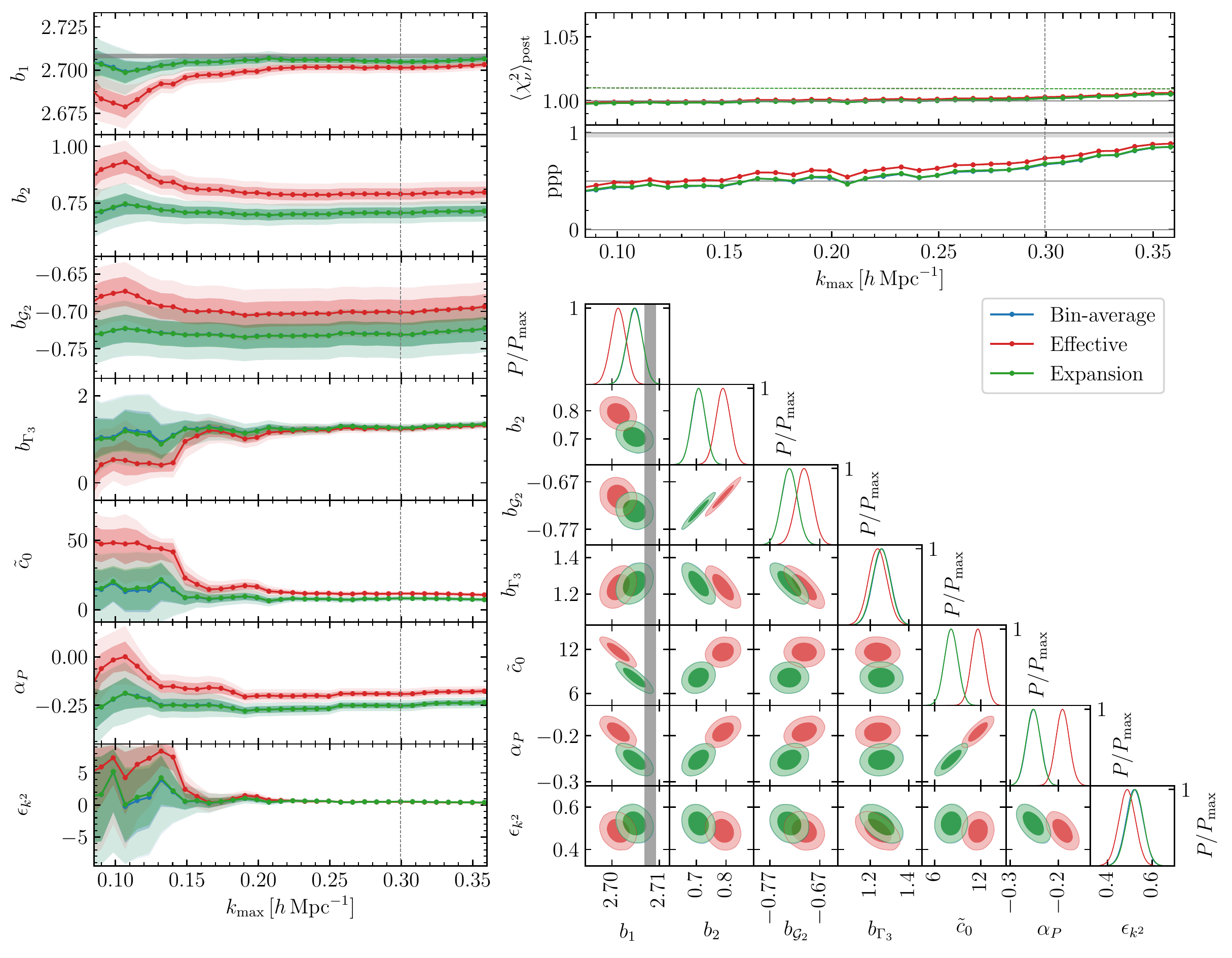}
    \caption{Same as figure \ref{fig:referencemodel}, but now comparing different methods to evaluate the theoretical models for the power spectrum and the bispectrum. In blue, the reference case with the bin-average of the theoretical prediction, in red the case where the models are evaluated at effective wavenumbers, and in green the case where the expansion method, described in appendix \ref{app:binning}, is used. Blue lines and contours coincide almost exactly with the green ones.}
    \label{fig:binning_comparison}
\end{figure}

The comparison, which considers the full volume of all \mine{} simulations, is shown in figure \ref{fig:binning_comparison}. Even in this rather challenging test, the results for the expansion method (in green) are essentially indistinguishable from the results assuming the exact binning (in blue, but exactly underneath the green areas). The effective approach shows instead some significant discrepancies: the posterior of the linear bias $b_1$ is clearly inconsistent with the value measured from the cross-to-matter ratio at all values of $\kmaxP$, and moreover significant tensions are present for most of the parameters, particularly at larger scales. The 2D marginalized posteriors also show these strong deviations, with the $b_2 - b_{\mathcal G_2}$ contours being completely inconsistent in the two cases.

These differences are evident because of the large volume considered and the corresponding small statistical uncertainties in both the power spectrum and the bispectrum. Still, such effects are typically larger for higher-order multipoles in redshift space and it is interesting to explore alternative, efficient methods to deal with Fourier-space discreteness. The expansion method allows us to compute an excellent approximation of the full bin-average of theoretical predictions in a time of the same order of magnitude needed for an evaluation with the effective method: we use it in section \ref{ssec:cosmores} to run the MCMC simulations where the parameter space includes cosmological parameters, since in this case the exact binning approach for the bispectrum would be impracticable.

%%%%%%%%%%%%%%%%%%%%%%%%%%%%%%%%%%%%%%%%%%%%%%%%
\subsection{Covariance approximations}
\label{ssec:covariance_app}

So far, all results assumed the covariance matrix for power spectrum and bispectrum, including the cross-covariance, estimated from the full set of 10\,000 \pin{} mocks. Such a large number of mocks is very often not available and it is necessary to resort to various approximations for the data covariance properties. We consider specifically three different cases in addition to our reference full covariance. In the first we exclude the cross-covariance between power spectrum and bispectrum, retaining the full individual covariance matrices for both statistics. The other two cases both consider the approximation that reduces the covariance matrix to its diagonal, requiring simply an estimate of the variance of power spectrum and bispectrum. In one case, this is estimated numerically from the mocks (and denoted as mock variance) while in the other we compute its Gaussian prediction from the power spectrum nonlinear, theoretical model assuming the bias parameters given by the best-fit values of the reference analysis (theoretical variance).

\begin{figure}
    \centering
    \includegraphics[width=\textwidth]{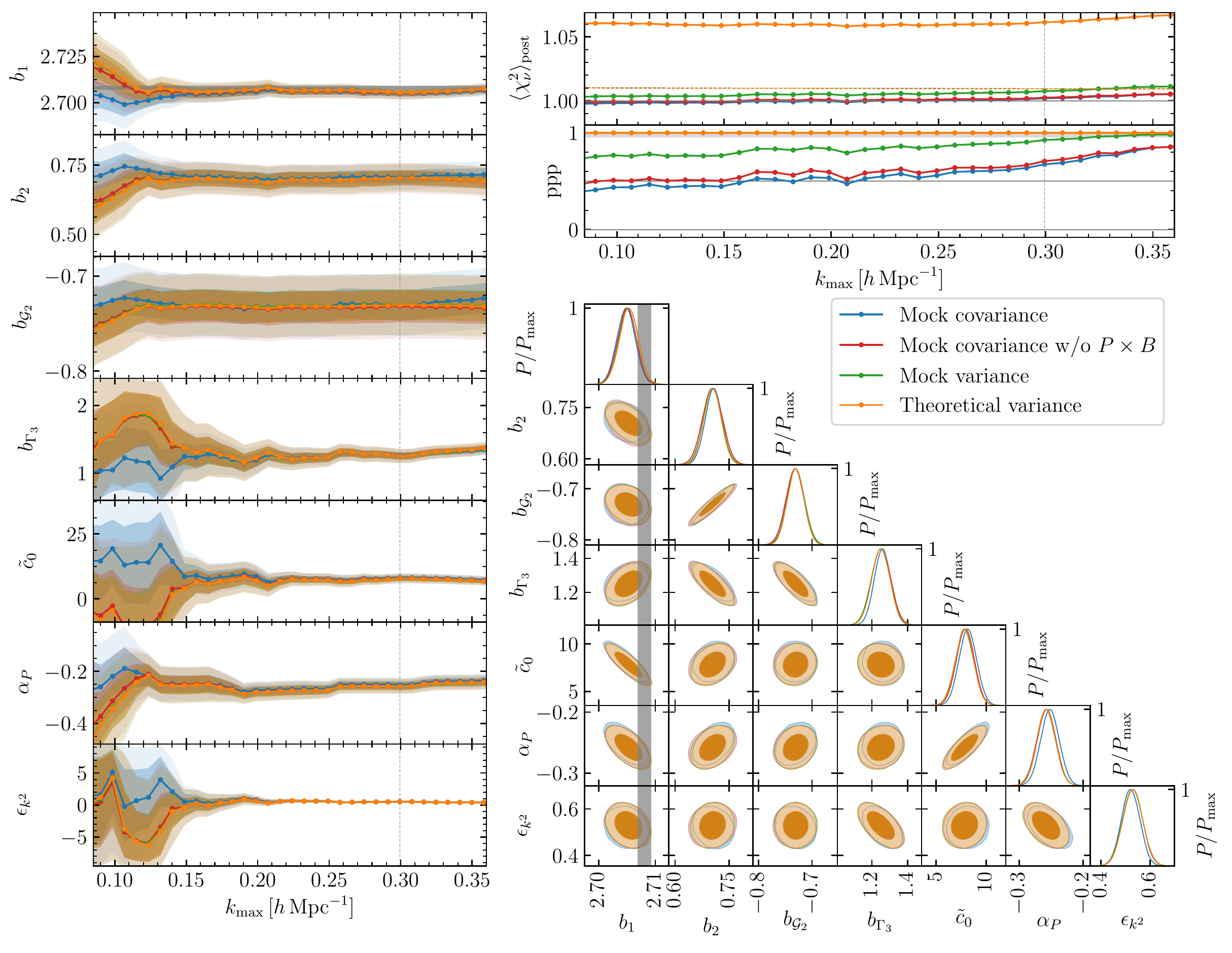}
    \caption{Same as figure \ref{fig:referencemodel}, but now comparing different approximation to the covariance matrix. In blue, the reference case with the full mock covariance; in red, the case where cross-correlations between power spectrum and bispectrum are set to zero; in green, the case where only the mock variance is used; in orange, the case where a theoretical Gaussian covariance is assumed.}
    \label{fig:cov_comparison}
\end{figure}

The comparison is shown in figure \ref{fig:cov_comparison}. All approximations are consistent with the reference case starting from the mildly non-linear regime, $\kmaxP \gtrsim 0.15 \kMpc$. At larger scales, we notice some differences between the reference case and all approximations, including the one excluding only the cross-covariance. This suggests that this contribution has some impact in the recovery of unbiased estimates of the model parameters, at least at the $1\sigma$ level. This is also consistent with the fact that these cross-correlations are expected to be quite large, with some of them being of the order of $40 \%$. We remind the reader that the single-parameter posteriors are shown as a function of $\kmax\equiv\kmaxP$, denoting the power spectrum range only, while for the bispectrum we fixed $\kmaxB=0.09 \kMpc$. It is possible that if both statistics covered the same range of scales these effects would appear at larger values of $k$.

Looking at goodness-of-fit metrics, the covariance approximations determined from the mocks, including the case of the sole variance, provide estimates of the range of validity of the model quite close to our reference case. The theoretical Gaussian variance, however, does not provide a good fit at any of the scales shown in the plot. Since the reduced chi-square is nearly constant as a function of $\kmax$, one possible explanation might be that the Gaussian approximation provides a bad estimate of the full bispectrum variance. Indeed, by direct comparison, we observe that the theoretical Gaussian variance is a few percent lower with respect to the \pin{} variance, and that differences reach $20\%$ for some squeezed triangles, even at relatively large scales. It is possible that for such configurations the non-Gaussian contributions can be particularly large, but we leave this to future investigations.

%%%%%%%%%%%%%%%%%%%%%%%%%%%%%%%%%%%%%%%%%%%%%%%%
\subsection{Triangle selection criteria}

In the power spectrum case, the range of validity of a given theoretical model is usually simply determined in terms of the largest wavenumber, $\kmaxP$, where the model provides a good fit to the data. In the bispectrum case, each value of $\kmaxB$ corresponds to a subset of triangles and we can expect a given theoretical model to perform more or less well on these configurations characterised by the same largest side but different shapes.

\begin{figure}
    \centering
    \includegraphics[width=\textwidth]{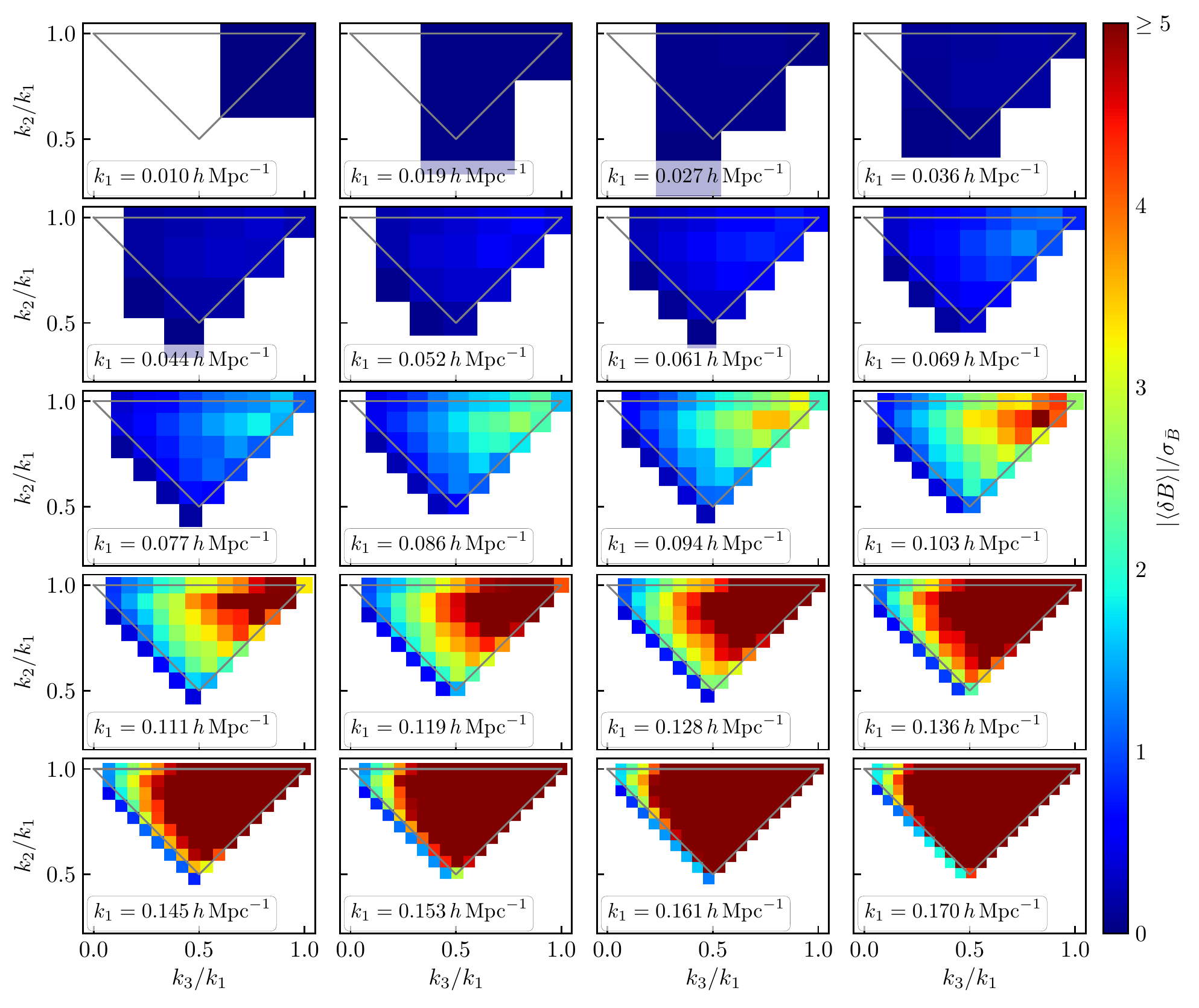}
    \caption{Mean residuals, normalized to the standard error on the mean, of all the measured bispectrum bins, assuming the posterior-averaged tree-level model of the reference analysis, $\kmaxP = 0.30 \kMpc$, $\kmaxB = 0.09 \kMpc$. In each panel, we fix the value of $k_1$ and show all triangle bins according to their value of $k_2$ and $k_3$. As a reference, the gray triangles in each panel represent the border of allowed fundamental triangles with $\vert \qv_1 \vert = k_1$. Tiles in the upper-right corner represent triangles closer to equilateral ($k_1 \simeq k_2 \simeq k_3$), those in the lower corner represent collinear triangles ($k_1 \simeq 2 k_2 \simeq 2 k_3$) while those in the upper-left corner represent squeezed triangles ($k_3 \ll k_1$). A bluer color shows a better agreement between model and data, while a redder color shows a worse fit. We cutoff the scale at $5\sigma$.}
    \label{fig:restriangles}
\end{figure}

This is illustrated in figure \ref{fig:restriangles}, where we show the mean residuals, normalized to the standard error on the mean, of all the measured bispectrum bins up to the maximum wavenumber $k_{1,\rm max} = 0.174 \kMpc$, assuming a posterior-averaged model from the reference joint fit. Each panel shows a subset of triangles characterised by the same value for the largest side $k_1$, shown as a function of the ratios $k_2/k_1$ and $k_3/k_1$ so that at the top-left we have squeezed configurations, at the top-right equilateral configurations, while at the bottom we have collinear, isosceles triangles ($k_1\simeq 2\,k_2 \simeq 2\,k_3$). 
So far we assumed for all our analysis a $\kmaxB = 0.09 \kMpc$. Including the whole subset of triangles with the next value of the largest side $k_1 > \kmaxB$ leads to the failure of the model to correctly describe the additional data, but we can expect this to happen first for nearly equilateral configurations, while a good fit can still be recovered for generic collinear configurations, i.e. with $k_1\simeq k_2+k_3$.

We want to check if different selection criteria based on other parameters than the sole $\kmaxB$ could lead to sensibly improved constraints on the model parameters. To this end we consider the following two triangle selection criteria:
\begin{itemize}
    \item we choose only triangles satisfying the condition $k_1 + k_3 \le \tilde k$ for a fixed value of $\tilde k$, and study parameter constraints as a function of $k_1$;
    \item we choose only triangles satisfying the condition $k_1 + k_2 \le \tilde k$ for a fixed value of $\tilde k$, and study parameter constraints as a function of $k_1$;
\end{itemize}
Both choices, that always assume $k_1\ge k_2\ge k_3$, allow to remove from the analysis nearly equilateral triangle bins, while keeping a subset of triangles with a different shape, as we include smaller scales by increasing $k_1$. This means that, when compared to the usual analysis, these selections lead to a smaller total number of triangles for the same value of $k_1=\kmaxB$, but, at the same time, they should provide a good fit for larger values of $\kmaxB$.

\begin{figure}
    \centering
    \includegraphics[width=\textwidth]{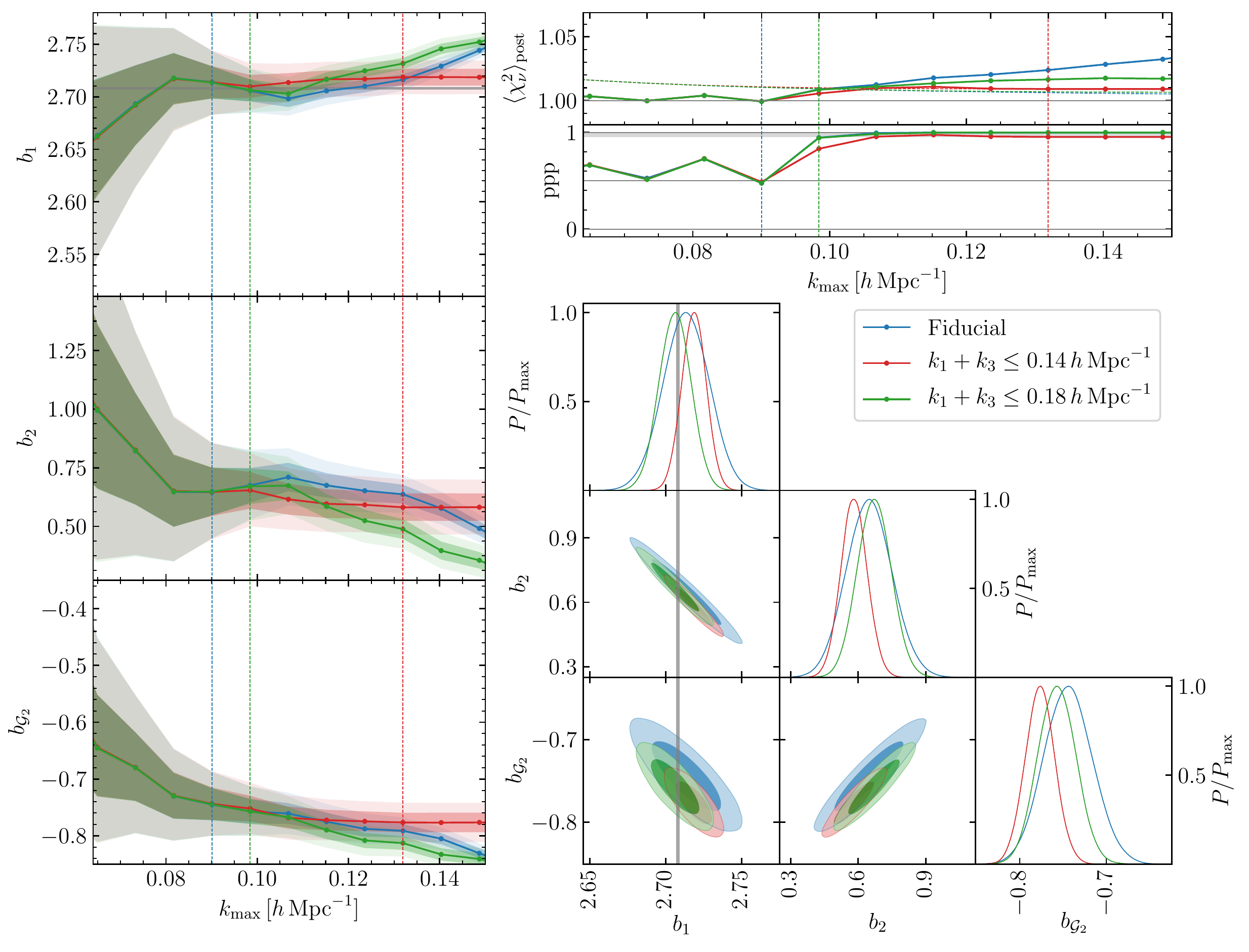}
    \caption{Comparison between the fits to the bispectrum-only data shown as a function of the largest wavenumber $\kmax$ included (blue) against the same analysis performed over data sets satisfying an additional condition of the maximum value of the combination $k_1+k_3$. We consider in particular $k_1+k_3\le 0.14\kMpc$ (red) and $k_1+k_3\le 0.18\kMpc$ (green). The 2D contour plots in this case correspond to different values of $\kmax$ defined as the largest values ensuring consistent results. They are marked with vertical lines of the corresponding colors in the left and top-right panels. In the contour plot, the number of triangles is 170, 222, 215 for the blue, red and green contours respectively.}
    \label{fig:k13}
\end{figure}

Figure \ref{fig:k13} shows (in blue) the results of the analysis of the bispectrum alone as a function of $\kmaxB$ as usually performed, compared with the same analysis where an additional condition is imposed to the combination $k_1+k_3$, reducing the total number of triangles. In particular we consider $k_1+k_3\le 0.14\kMpc$ (red) and $k_1+k_3\le 0.18\kMpc$ (green). We notice that the stricter condition $k_1+k_3\le 0.14\kMpc$ allows the inclusion of only a few more configurations w.r.t. those included in the standard result for $\kmax=0.09\kMpc$. In fact, for $\kmax>0.13\kMpc$ the posteriors do not change as no additional configuration can satisfy the condition and the quality of the fit remains acceptable for all the selected triangles. Comparing these results at $\kmax=0.14\kMpc$ with the usual ones at $\kmax=0.09 \kMpc$ we find a non-negligible improvement on the parameters constraints of almost 50\%. Imposing the condition on the sum $k_1+k_3$ with the larger value $k_1+k_3\le 0.18\kMpc$ allows for too many triangles not properly described by the model to be included, leading quickly to significant systematic errors on the recovered parameters. 

\begin{figure}
    \centering
    \includegraphics[width=\textwidth]{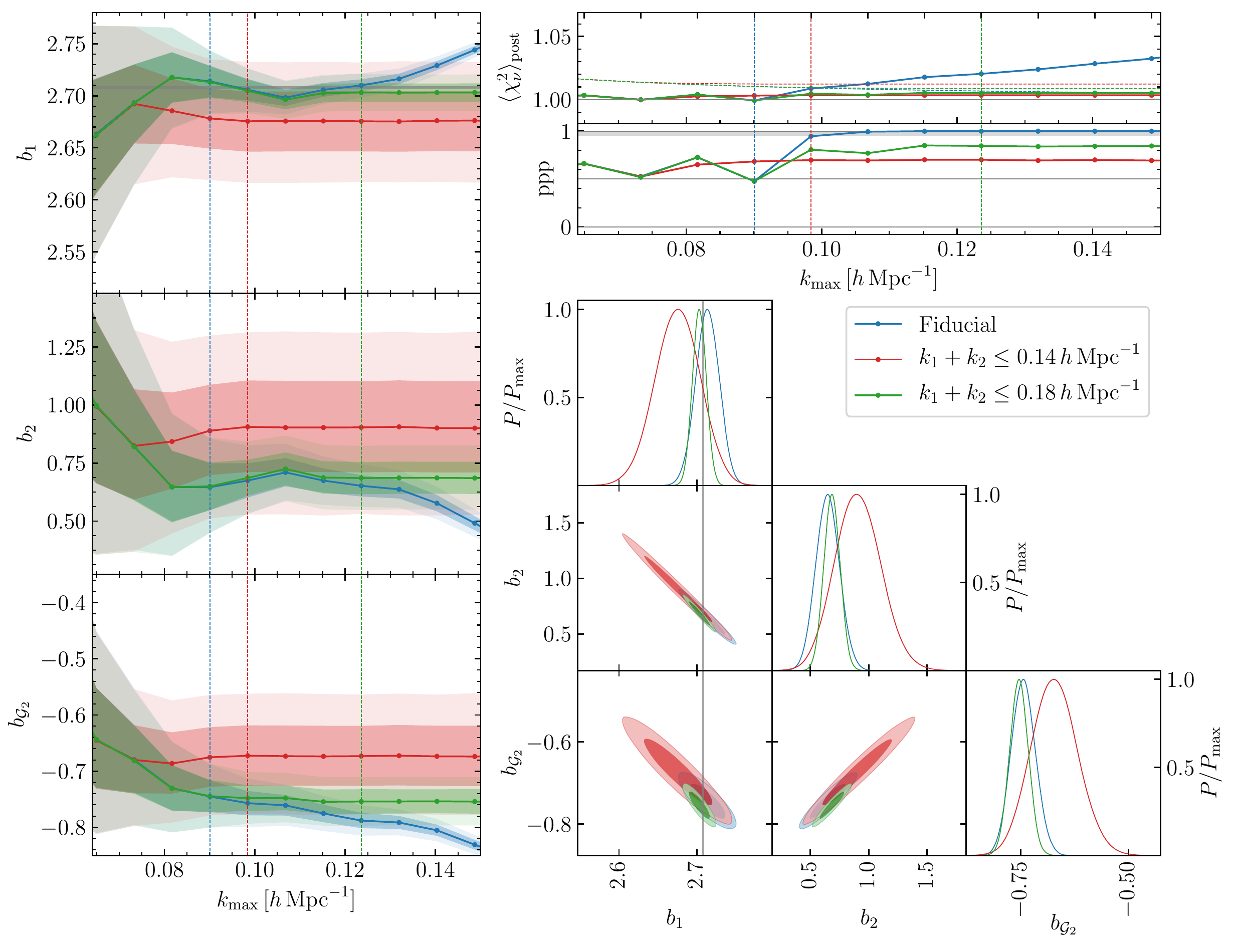}
    \caption{Same as fig.~\ref{fig:k13} but with the additional condition on the combination $k_1+k_2$. We consider $k_1+k_2\le 0.14\kMpc$ (red) and $k_1+k_2\le 0.18\kMpc$ (green). This time, in the contour plot, the number of triangles is 170, 121, 236 for the blue, red and green contours respectively.}
    \label{fig:k12}
\end{figure}

In figure \ref{fig:k12} we consider instead a condition on the sum $k_1+k_2$. In this case, both $k_1+k_2<0.14\kMpc$ and $k_1+k_2<0.18\kMpc$ do not allow for any additional configurations for $\kmax>0.12\kMpc$ and in both cases we retrieve constraints on $b_1$ consistent with the expected value $b_1^\times$. However, only the looser condition $k_1+k_2<0.18\kMpc$ provides better constraints, of about 30\%, than the standard case with $\kmax=0.09\kMpc$.

We expect the introduction of selection criteria of this kind to be particularly relevant for constraining non-Gaussian initial conditions of the local type, where the signal peaks in the squeezed configurations \cite{SefusattiCrocceDesjacques2012}. We limit ourselves to remark, here, that the improvement, even when small, is obtained at no additional cost.

%%%%%%%%%%%%%%%%%%%%%%%%%%%%%%%%%%%%%%%%%%%%%%%%
\subsection{Inference of cosmological parameters}
\label{ssec:cosmores}

We finally present the results for the joint fit of the halo power spectrum and bispectrum in real space aimed at recovering unbiased estimates of the cosmological parameters. We use here the total volume of the \mine{} simulations. The theoretical model coincides with the reference model depending on the seven bias and shot-noise parameters in eq.~(\ref{e:ref_params}). In this case, however, also the three cosmological parameters $A_s$, $h$, and $\omega_m$ are let free to vary within a range specified by uniform priors (see table~\ref{tab:priorsPB}). Notice that, in practice, the bias parameters that we vary are the ones defined in eq.~(\ref{e:A_params}) (but we show the original ones in the figure). This alleviates the strong degeneracy between {\eg} $b_1$ and $A_s$, and thus speeds up the convergence of the MCMC runs. All parameters are varied consistently at each step of the MCMC runs, the theoretical model is recomputed fully, and Fourier-grid effects are accounted for by means of the expansion method described in appendix \ref{app:binning}. The covariance matrix is again the one estimated from the \pin{} mocks.

\begin{figure}
    \centering
    \includegraphics[width=\textwidth]{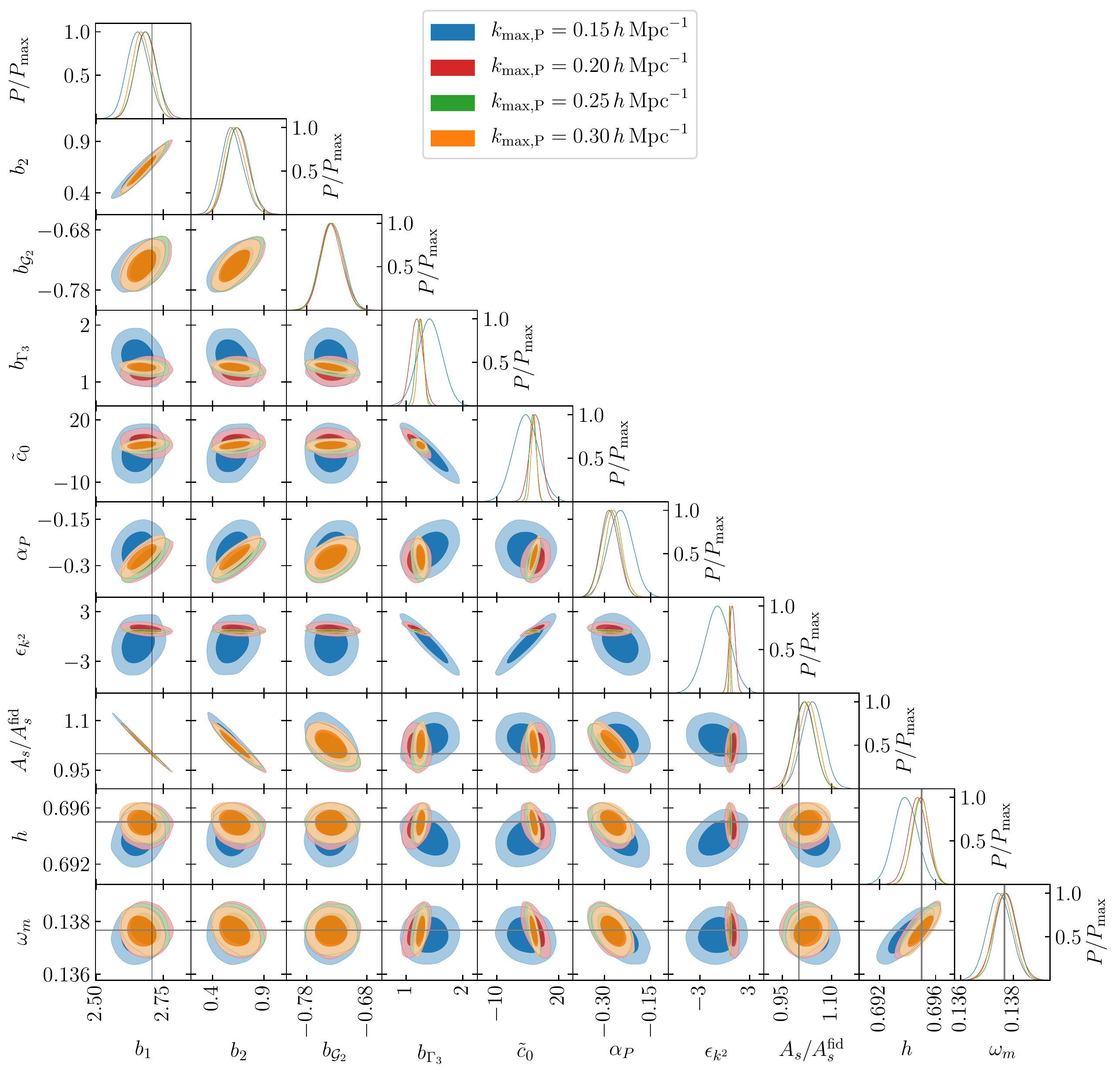}
    \caption{Triangle plot showing the 1D and 2D marginalized posteriors of the parameters of the fiducial model, where also three cosmological parameters are varied, from a joint fit of the halo power spectrum and bispectrum measured from the N-body simulations, for different values of $\kmaxP = 0.15, 0.20, 0.25, 0.30 \kMpc$ (in blue, red, green, and orange respectively). Gray lines show the linear bias measured from the cross power spectrum and the input values of the cosmological parameters.}
    \label{fig:cosmo_PB}
\end{figure}

Figure \ref{fig:cosmo_PB} shows the parameter constraints on the model parameters for four different values of $\kmaxP = 0.15, 0.20, 0.25, 0.30 \kMpc$ (for the bispectrum, we still set $\kmaxB = 0.09 \kMpc$). The posteriors for the cosmological parameters are nicely consistent with the input values used to run the N-body simulations, marked with gray lines, and the one for the linear bias $b_1$ again agrees with the value measured from the cross halo-matter power spectrum. This agreement is clearly visible in the 2D marginalized posteriors as well. Moreover, all the other parameters are still perfectly consistent with the values extracted from the reference analysis with fixed cosmological parameters. It is also worth noticing how, regardless of the degeneracies that might be present between bias and cosmological parameters, the posteriors for the bias parameters are still stable as a function of $\kmaxP$. Similar conclusions can also be drawn for the parameter $\tilde c_0$ and for the stochastic parameters.

\begin{figure}
    \centering
    \includegraphics[width=0.70\textwidth]{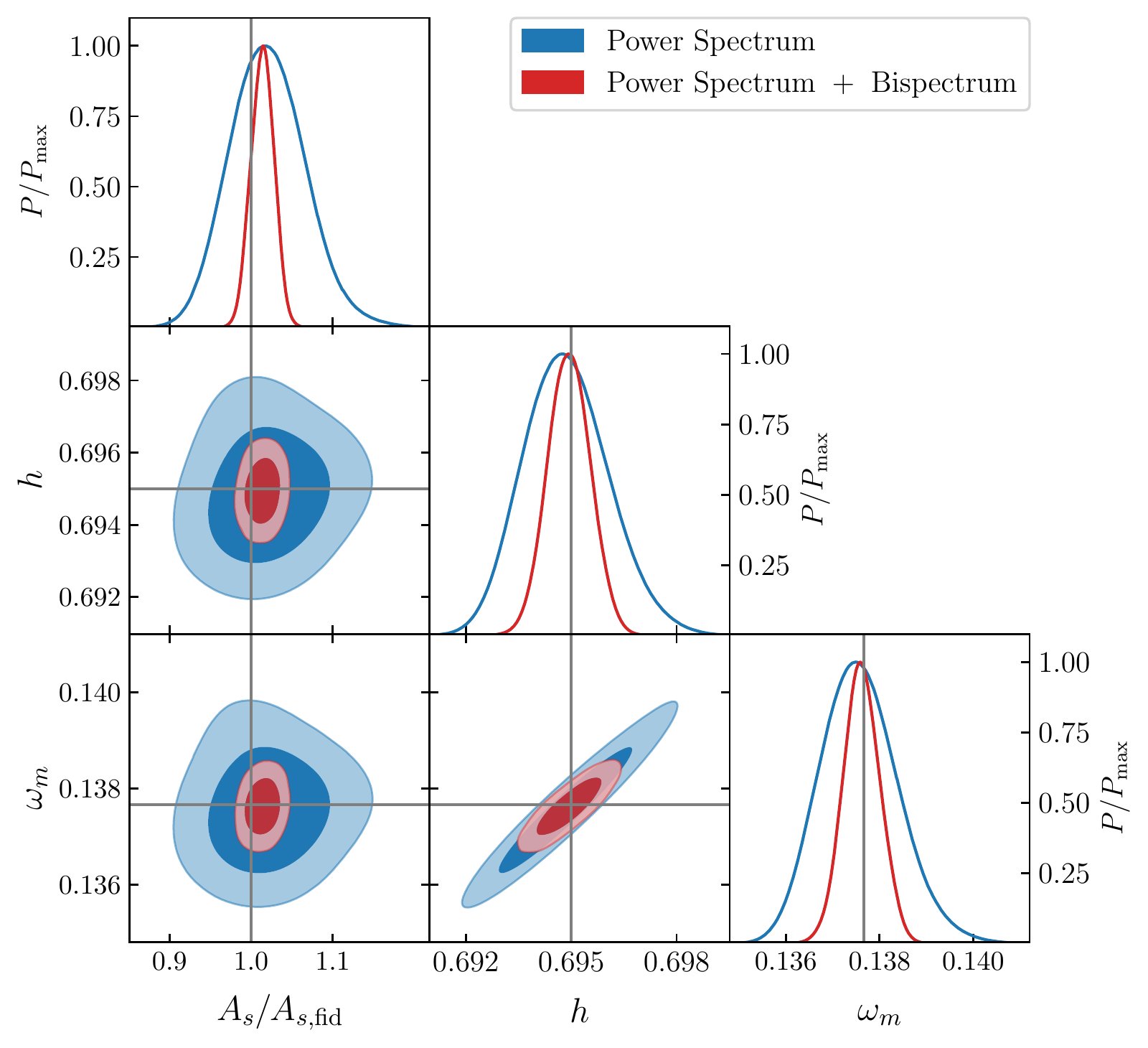}
    \caption{Triangle plot showing the 1D and 2D marginalized posteriors of the cosmological parameters inferred through a likelihood analysis of power spectrum (blue) and power spectrum and bispectrum (red) using the fiducial model. In this plot, we marginalize over all bias and stochastic parameters in order to highlight the impact that the inclusion of the bispectrum to the analysis has on the inference of cosmological parameters. For the power spectrum we set $\kmaxP = 0.30 \kMpc$, while we set $\kmaxB = 0.09 \kMpc$ for the bispectrum. Gray lines show the input values of the cosmological parameters.}
    \label{fig:cosmo_P_PB}
\end{figure}

In figure \ref{fig:cosmo_P_PB}, we compare the constraints obtained fitting the power spectrum only (up to $\kmaxP = 0.30 \kMpc$) and performing a joint fit of the power spectrum and the bispectrum (for the latter we use $\kmaxB = 0.09 \kMpc$). The fiducial model discussed in section \ref{ssec:reference} is fitted to the data but, once again, we let the cosmological parameters $A_s$, $h$, and $\omega_m$ vary. Strong parameter degeneracies are present in the power spectrum-only fit and the Markov chains do not satisfy the formal convergence criterion we use even after very many steps. Therefore, it is possible that the size of the blue constraints in the figure are underestimated, although we suspect not by much (based on multiple MCMC chains). The inclusion of the bispectrum to the analysis, even at large scales, tightens the constraints on the cosmological parameters: constraints on $A_s$ are reduced by a factor $3.3$, while the ones on $h$ and $\omega_m$ by a factor of roughly $2$.

The data set we have analysed does not capture the full complexity of a galaxy redshift survey. Our setup, based on simulations within periodic boxes, at fixed redshift, and in real space, still lacks a proper modelling (in the observables, and possibly in the covariances) of non-linearities arising from redshift-space distortions, and of mode-coupling effects due to the survey window function. To an extent, this might result in a minor improvement on the constraints of the model parameters when the bispectrum is included in a realistic data analysis. However, given the unprecedentedly large volume we considered (1000$\cGpc$, almost twice the volume analysed in the challenge paper of \cite{NishimichiEtal2020}), our results form a very stringent test of halo bias models, as well as a strong consistency check between perturbative models and the non-linear dynamics simulated by N-body solvers. They also provide strong evidence that the perturbative bias treatment and the counterterms do not distort the posterior distribution of the cosmological parameters, at least in real-space. We thus conclude that a joint likelihood analysis of the power spectrum and the bispectrum should be able to provide unbiased estimates for the cosmological parameters, including information on the accelerated expansion of the Universe.

%%%%%%%%%%%%%%%%%%%%%%%%%%%%%%%%%%%%%%%%%%%%%%%%
%%%%%%%%%%%%%%%%%%%%%%%%%%%%%%%%%%%%%%%%%%%%%%%%
\section{Conclusions}
\label{sec:conclusions}

We presented a joint likelihood analysis of the real-space halo power spectrum and bispectrum extracted from 298 N-body simulations covering a total volume of roughly $1000 \cGpc$. We compared the data to a perturbative model at one-loop for the power spectrum and at tree-level for the bispectrum. The model implementation, limited here to real space, is essentially the same that has been recently applied to the analysis of the BOSS data in \cite{IvanovSimonovicZaldarriaga2020}. 
In order to estimate the full non-linear covariance matrix for both observables along with their cross-covariance, we used measurements from 10\,000 mock halo catalogs generated with the \pin{} code. We can summarize the main results of our analysis as follows.
\begin{itemize}
    \item Using Bayesian model selection, we identify the optimal set of free parameters that can be constrained by the data (with a fixed background cosmological model), namely four bias parameters, one counterterm parameter, and two stochastic corrections to the power spectrum model. 
    \item The theoretical model for the power spectrum nicely fits our numerical data up to $\kmaxP \sim 0.3 \kMpc$. Considering the power spectrum along with the bispectrum (up to $\kmaxB = 0.09 \kMpc$), the fit provides unbiased estimates of the linear bias parameter $b_1$ with sub-percent precision, as well as a good fit to the data as estimated both in terms of the $\chi^2$ and $\rm ppp$ diagnostics -- even when the full data set is considered.
    \item We explore the possibility of reducing the dimensionality of parameter space by assuming that not all the bias parameters are independent, as suggested by several numerical and theoretical studies. In all cases, fitting the simplified models to the full data set gives biased estimates of the parameters. However, when the probed volume is reduced to match those that will be covered by the upcoming surveys ($6\cGpc$), all the fits based on the bias relations provide consistent values of the free parameters with smaller uncertainties than the default case. In particular, the DIC indicates that the data are best described by the $\bG(b_1)$ relation proposed by \cite{EggemeierEtal2020}.
    \item We investigate different methods to account for the discrete nature of measurements of Fourier-space correlators in the binning of the theoretical predictions. We find that, when the larger volume is considered, the evaluation of the model at a single effective triangle per bin leads to strongly biased parameter constraints. We propose a new method, discussed in appendix \ref{app:binning}, and we show that it is able to provide constraints consistent with our reference results (obtained by averaging exactly the model over the Fourier wavenumbers in each bin).
    \item We test several approximations to the covariance matrix. We find that neglecting the cross-correlations between power spectrum and bispectrum slightly biases the constraints on the model parameters. In addition, we show that the Gaussian (diagonal) approximation to the covariance matrix underestimates the errors by up to 20\% on some triangular configurations, and fails to provide a proper estimate of the goodness-of-fit of the theoretical model.
    \item We explore different selection criteria to reduce the number of triangular configurations for the analysis of the bispectrum. We find that a selection of the triangular configurations accounting as well for their shape, rather than only the largest wavenumber $\kmaxB$, can lead to an improvement in the parameters constraints by up to 50\%.
    \item Finally, we perform a likelihood analysis in which also three cosmological parameters are varied. In this case, we use the power spectrum and the bispectrum data extracted from the full simulation suite. The constraints on the cosmological parameters obtained with our default 7-parameter model are nicely consistent with the input values of the simulations, up to $\kmaxP = 0.3 \kMpc$. Moreover, compared to a power spectrum-only analysis, the constraints on cosmological parameters in a joint analysis shrink significantly, by a factor of $\sim 3$ for the amplitude of scalar perturbations $A_s$ and by a factor of $\sim 2$ for the Hubble parameter $h$ and the relative abundance of matter $\omega_m$. This major achievement demonstrates the feasibility of using perturbative models with free parameters in order to extract information on the underlying cosmological parameters from the joint analysis of the power spectrum and the bispectrum.
\end{itemize}

As already mentioned in the introduction, the likelihood analysis of three-point statistics is still a relatively poorly explored subject (particularly in order to set constraints on the cosmological parameters). While the ideal data set considered here does not have the complexity of a galaxy redshift survey, its large total simulation volume (combined with the 10,000 mock catalogs), allowed us to investigate the impact of several assumptions which are routinely made in this kind of studies. Upcoming observations will require a better quantification and control over possible systematic errors both in the theoretical modelling as in the methodology. For these reasons, we think that our work is a step towards more rigorous and thorough analysis of spectroscopic redshift surveys.

\acknowledgments

We are particularly grateful to Claudio Dalla Vecchia and Ariel Sanchez for running and making available the \mine{} simulations set, performed and analysed on the Hydra and Euclid clusters at the Max Planck Computing and Data Facility (MPCDF) in Garching.
The \pin{} mocks were run on the GALILEO cluster at CINECA, thanks to an agreement with the University of Trieste.
We acknowledge useful discussions with Martin Crocce, Alex Eggemeier, Azadeh Moradinezhad, Chiara Moretti, Andrea Pezzotta, Roman Scoccimarro, Zvonimir Vlah. 
We acknowledge the hospitality of the Institute for Fundamental Physics of the Universe in Trieste where part of this work was carried out in October 2019.
ES and PM are partially supported by the INFN INDARK PD51 grant and acknowledge support from PRIN MIUR 2015 Cosmology and Fundamental Physics: illuminating the Dark Universe with Euclid.
AO thanks the Institute of Space Sciences (IEEC-CSIC) in Barcelona, where part of this work has been developed, for hospitality and support, and the Erasmus+ program that made this possible.

\appendix
%%%%%%%%%%%%%%%%%%%%%%%%%%%%%%%%%%%%%%%%%%%%%%%%%%%%%%%%%
%%%%%%%%%%%%%%%%%%%%%%%%%%%%%%%%%%%%%%%%%%%%%%%%%%%%%%%%%

\section{Approximations for the bin-averaged theoretical predictions}
\label{app:binning}

%%%%%%%%%%%%%%%%%%%%%%%%%%%%%%%%%%%%%%%%%%%%%%%%%%%%%%%%%
\subsection{Power spectrum}
Within a Fourier bin, the theoretical model for the power spectrum can be expanded in Taylor series around some wavenumber $q_0$ included in the same bin as
\be
P(q) = \sum_{n = 0} ^{\infty} \frac{1}{n!}P^{(n)}(q_0)(q-q_0)^n,
\ee
so that its bin-average can be written as
\be
    P_{\rm bin}(k) = \frac{1}{N_P(k)}\sum_{\qv \in k}\sum_{n = 0} ^{\infty} \frac{1}{n!}P^{(n)}(q_0)(q-q_0)^n = \frac{1}{N_P(k)}\sum_{n = 0} ^{\infty}\frac{1}{n!}P^{(n)}(q_0)\sum_{\qv \in k} (q-q_0)^n\,.
\ee
By defining the quantities
\be
    \mu_{n}(k) =\frac{1}{N_P(k)}\sum_{\qv \in k} (q-q_0)^n,
\ee
we can write the bin-average of the power spectrum as
\be
    P_{\rm bin}(k) = \sum_{n = 0} ^{\infty}\frac{1}{n!}P^{(n)}(q_0)\mu_{n}(k)\,.
    \label{eq:Pfullexp}
\ee
This expression does not involve the evaluation of the power spectrum and its derivatives at each value of $q=|\qv|$, and the quantities $\mu_n(k)$ can be pre-computed. 

If we choose for $q_0$ the effective wavenumber $k_{\rm eff}$ defined in eq.~(\ref{eq:keffP}), the generic $\mu_n(k)$ reduces to the central $n$-th moment of the discrete distribution of Fourier wavenumbers in the bin. The zero-th order term in the infinite expansion eq.~(\ref{eq:Pfullexp}), reduces to the standard effective power spectrum of eq.~(\ref{eq:Peff}), since $\mu_0(k) = 1$. The first order contribution vanishes since $\mu_1(k) = 0$.
The first, non-vanishing correction to the zero-th order term is then given by the second-derivative term, with $\mu_2(k)$ being the variance of the distribution of Fourier wavenumbers inside the bin. Therefore, we can approximate the bin-averaged power spectrum, truncating the expansion to include up to the second order, as
\be
P_{\rm bin}^{(2)}(k) \simeq P(\keff) + \frac{1}{2}P^{\prime \prime}(\keff) \mu_2(k).
    \label{eq:PTeffPreal}
\ee

\begin{figure}
    \centering
    \includegraphics[width=\textwidth]{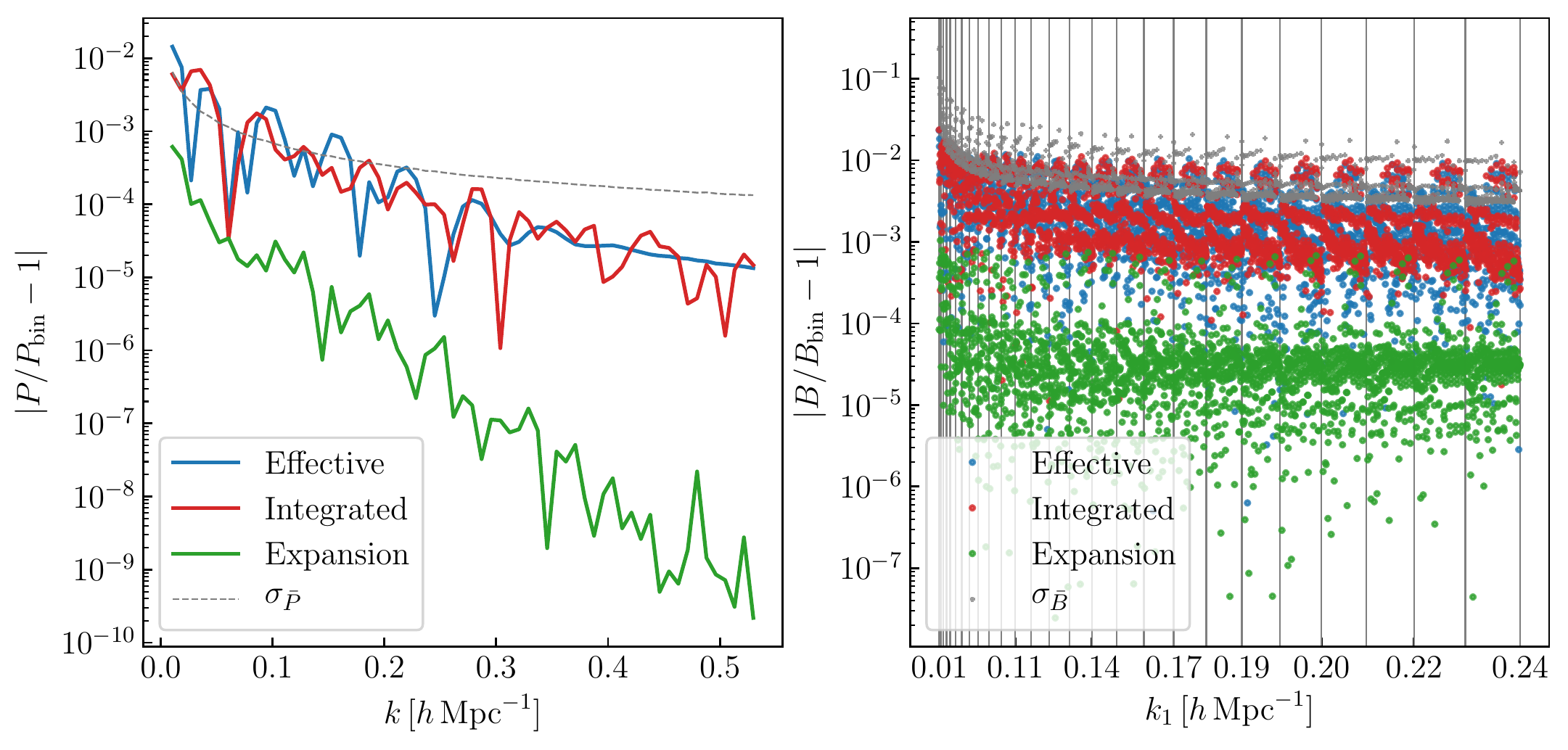}
    \caption{Left panel: relative difference between the bin-averaged theoretical model of the halo power spectrum and the model evaluated at the effective wavenumbers $\keff$ (in blue), the model evaluated using eq. (\ref{eq:PTeffPreal}) based on a Taylor expansion about $\keff$ (in green), and the model averaged approximating the discrete sums with continuous integrals (in red); in gray, the relative standard error on the mean (estimated from the mocks) of the full dataset of 298 N-body simulations. Right panel: same comparison for the bispectrum case.}
    \label{fig:expcomparison}
\end{figure}

In order to have a well-behaved, continuous $n$-th derivative, it is required that the starting power spectrum is interpolated with at least an $(n+1)$-th order spline. This means that a cubic spline interpolation of the theoretical model of the power spectrum is sufficient for our purposes. The quantities $\keff$ and $\mu_2(k)$ are evaluated only once, and therefore the computational cost of this approach is of the same order of the usual effective approach, while providing a great improvement in accuracy. This is shown in the left panel of figure \ref{fig:expcomparison}, where the relative difference with the fully bin-averaged power spectrum is compared with the one of the effective prediction. As a reference, we also compare it with the integral approximation of the exact bin-average, and with the relative statistical uncertainty of our dataset.

%%%%%%%%%%%%%%%%%%%%%%%%%%%%%%%%%%%%%%%%%%%%%%%%%%%%%%%%%
\subsection{Bispectrum}

In the case of the bispectrum, due to its shape-dependence, the Taylor expansion approach is model dependent. For this reason, in the following it is more convenient to assume explicitly the structure of the tree-level model, eq.~(\ref{eq:Bspt}), that we adopted for the galaxy bispectrum. Introducing the generic kernel $\widetilde{K}$, we can write the model in the form
\be
    B(\qv_1, \qv_2, \qv_3) = \Kt(\qv_1, \qv_2)\PL(q_1)\PL(q_2) + {\rm cyc.}\,.
\ee
We also make use of the following notation for averages over the triangular bin with sides $(k_1,k_2,k_3)$
\be
\langle f(\qv_1, \qv_2, \qv_3) \rangle_{\bigtriangleup}\equiv \frac{1}{N_B(k_1, k_2, k_3)}\sum_{\qv_1 \in k_1} \sum_{\qv_2 \in k_2} \sum_{\qv_3 \in k_3} \delta_K(\qv_{123}) f(\qv_1, \qv_2, \qv_3) \,.
\ee
Then, inside the triangle bin, we can expand the product of the power spectra in Taylor series around the sorted effective wavenumbers defined in eq.s~(\ref{eq:keffB})
\begin{align}
    \PL(q_l)\PL(q_m) &= \sum_{u = 0}^{\infty}\sum_{v = 0}^{\infty}\frac{1}{u!\, v!}\PL^{(u)}(\keffl)\PL^{(v)}(\keffm)(q_l-\keffl)^u(q_m-\keffm)^v,
\end{align}
where $(\qv_1, \qv_2, \qv_3)$ are relabeled as $(\qv_l, \qv_m, \qv_s)$ (ordered from the longest to the shortest), and the full expression for the bin-average of the bispectrum model becomes
\be
    B_{\rm bin}(k_1, k_2, k_3) = \sum_{u = 0}^{\infty}\sum_{v = 0}^{\infty}\frac{1}{u!\, v!}\PL^{(u)}(\keffl)\PL^{(v)}(\keffm)\left \langle \Kt(\qv_l, \qv_m)(q_l-\keffl)^u(q_m-\keffm)^v \right \rangle_\bigtriangleup + {\rm cyc.}
    \label{eq:Bfullexp}
\ee
At zero-th order, we have that the approximation to the full bin-average of the bispectrum is simply
\be
    B_{\rm bin}^{(0)}(k_1, k_2, k_3) \simeq \left \langle \Kt(\qv_l, \qv_m) \right \rangle_{\bigtriangleup}\PL(\keffl)\PL(\keffm) + {\rm cyc.}\,.
    \label{eq:PTeffBrealLO}
\ee
Notice that this expression does not reduce to the bispectrum evaluated at effective wavenumbers, since it includes the exact bin-average of the kernel. At higher order, new terms appear, where the bin-average now applies to the product of the kernel with powers of Fourier wavenumbers. Truncating the full expansion, retaining terms up to $u+v = 2$, this approximation requires to precompute 18 averages for each kernel, in addition to the three effective wavenumbers.

The shot-noise contribution does not require the computation of any extra term, since the averages appearing are already computed for the constant kernel relative to the quadratic bias operator, 
\be
    B_{\rm shot-noise}^{\rm bin}(k_1, k_2, k_3) \simeq \frac{1+\alpha_1}{\bar n}b_1^2 \left[ \PL(\keffl) + \frac{1}{2} \PL''(\keffl) \left \langle \left(q_l - \keffl \right)^2\right \rangle_\bigtriangleup  \right] + {\rm cyc.} + \frac{1+\alpha_2}{\bar n^2}\,.
\ee

As shown in the right panel of figure \ref{fig:expcomparison}, this method provides generally a better accuracy (of at least one order of magnitude) compared to the standard effective method with sorted wavenumbers and to the integral approximation of the exact bin-average.

\setlength{\bibsep}{2pt plus 0.5ex}
\bibliographystyle{JHEP}
\bibliography{cosmologia}

\end{document}